\documentclass[fleqn,usenatbib]{mnras}

\usepackage{newtxtext,newtxmath}

\usepackage[T1]{fontenc}
\usepackage{ae,aecompl}

\usepackage{graphicx}	
\usepackage{amsmath}	
\usepackage{amssymb}	

\title[Decoupling the rotation of stars and gas - I]{Decoupling the rotation of stars and gas - I: the relationship with morphology and halo spin}

\author[C. Duckworth et al.]{Christopher Duckworth,$^{1,2}$\thanks{E-mail: cd201@st-andrews.ac.uk}
Rita Tojeiro,$^{1}$
Katarina Kraljic$^{3}$
\\
{}$^{1}$School of Physics and Astronomy, University of St Andrews, North Haugh, St Andrews, KY16 9SS, UK\\
$^{2}$Center for Computational Astrophysics, Flatiron Institute, 162 Fifth Avenue, New York, NY 10010, USA\\
{}$^{3}$Institute for Astronomy, University of Edinburgh, Royal Observatory, Blackford Hill, Edinburgh EH9 3HJ, UK\\
}

\date{Accepted XXX. Received YYY; in original form ZZZ}

\pubyear{2019}

\begin{document}
\label{firstpage}
\pagerange{\pageref{firstpage}--\pageref{lastpage}}
\maketitle

\begin{abstract}
We use a combination of data from the MaNGA survey and MaNGA-like observations in IllustrisTNG100 to determine the prevalence of misalignment between the rotational axes of stars and gas. 
This census paper outlines the typical characteristics of misaligned galaxies in both observations and simulations to determine their fundamental relationship with morphology and angular momentum. We present a sample of $\sim 4500$ galaxies from MaNGA with kinematic classifications which we use to demonstrate that the prevalence of misalignment is strongly dependent on morphology. The misaligned fraction sharply increases going to earlier morphologies (28$\pm$3\% of 301 early-type galaxies, 10$\pm$1\% of 677 lenticulars and 5.4$\pm$0.6\% of 1634 pure late-type galaxies). For early-types, aligned galaxies are less massive than the misaligned sample whereas this trend reverses for lenticulars and pure late-types. 
We also find that decoupling depends on group membership for early-types with centrals more likely to be decoupled than satellites. 
We demonstrate that misaligned galaxies have similar stellar angular momentum to galaxies without gas rotation, much lower than aligned galaxies. Misaligned galaxies also have a lower gas mass than the aligned, indicative that gas loss is a crucial step in decoupling star-gas rotation. 
Through comparison to a mock MaNGA sample, we find that the strong trends with morphology and angular momentum hold true in IllustrisTNG100. We demonstrate that the lowered angular momentum is, however, not a transient property and that the likelihood of star-gas misalignment at $z = 0$ is correlated with the spin of the dark matter halo going back to $z = 1$. 
\end{abstract}

\section{Introduction}
Angular momentum is one of the key properties that quantifies a galaxy. Within the $\Lambda$ cold dark matter ($\Lambda$CDM) paradigm, galaxies form from the cooling and condensation of the initial gas cloud within dark matter haloes \citep{fall1980, mo1998}. In this framework, the angular momentum content of the collapsing baryons are inherited from that of the surrounding dark matter halo \citep[tidal torque theory (TTT); e.g.][]{hoyle1951, peebles1969, Doroshkevich1970}. This is a natural result of the baryons and dark matter being well mixed at early times leading them to experience similar torquing from the surrounding tidal field of protohaloes. 

If gravitational collapse proceeds unhindered, the initial gas cloud will form a stable rotating disc which eventually evolves into the late type galaxies (LTGs) we observe today \citep{white1978}. Since stars form from the rotating gas, the natural expectation is that they will inherit its dynamical characteristics leading to coherent rotation between dark matter, gas and stars. 

The evolution of a galaxy from initial collapse to today is, however, seldom completed in isolation. By its very nature, structure formation in $\Lambda$CDM is hierarchical with haloes undergoing bottom-up assembly from mergers of lower mass progenitors. After turnaround, the angular momentum of the baryons in a galaxy can be driven dramatically away from the expectations of TTT through external processes such as interactions or mergers. How such interactions alter angular momentum depend on the magnitude, orientation and gas content of the merger. For example, gas rich mergers in general spin up galaxies whereas gas poor mergers are seen to spin down galaxies \citep[][]{lagos2017,lagos2018}.


Developments in spectrographs have led to the advent of integral field spectroscopy (IFS) which provides spatially resolved spectra for galaxies. Establishing work in the field has been the Spectrographic Areal Unit for Research on Optical Nebulae \citep[SAURON;][]{sauron} and ATLAS\textsuperscript{3D} \citep{atlas3d} surveys, which have focused on early type galaxies (ETGs) in the local Universe. IFS has enabled kinematic classification through a proxy for angular momentum based on the stellar kinematics up to one effective radius ($R_e$). Termed $\lambda_{Re}$, the measure enabled the clear distinction between slow and fast rotating ETGs \citep{emsellem2007, emsellem2011}. While there is still debate over whether 1$R_{e}$ is large enough to fully encapsulate the complete kinematic morphology of a galaxy \citep{foster2013,arnold2014}, it has opened the door for understanding the relationship between optical morphology and angular momentum. 

IFS surveys for $\sim$1000 of galaxies across all optical morphologies are now a reality. For example, the Sydney-AAO  Multi-object  Integral  field  spectrograph  survey \citep[][]{croom2012, bryant2015} has mapped $\sim$3400 galaxies upto $z\sim0.12$ across a variety of environments. Even larger is the Mapping Nearby Galaxies at Apache Point \citep[MaNGA;][]{bundy2015, blanton2017} survey which will map $\sim$10000 galaxies in the local Universe ($z=0-0.15$). By design MaNGA will create a sample of near flat number density distribution across absolute $i$-band magnitude and stellar mass.

Recent studies in these surveys and also simulations have demonstrated the close interlink between stellar angular momentum, stellar mass and morphology suggesting that late types and early type fast rotators form a continuous sequence rather than from fundamentally different formation pathways \citep[][]{cortese2016, lagos2017, graham2018}. Remarkably, despite the highly non-linear processes involved, current cosmological surveys predict that the stellar angular momentum in rotationally supported galaxies at $z=0$ is still conserved from that of the dark matter halo \citep[e.g.][]{genel2015}. 

In the extended theory of TTT, the spin of galaxies embedded in the larger-scale environment of the cosmic web can be seen to align with the direction of filaments \citep[e.g.][]{pichon2011,codis2015, laigle2015}. Low mass disks can accrete material most efficiently when its spin vector is aligned with the direction along the filament. Conversely, higher mass haloes can be formed through mergers in the plane along the filament, leading to a perpendicular spin alignment with the large scale structure. 

In this framework, it is then perhaps surprising to identify galaxies with decoupled rotation between the stars and gas. The ability of a given galaxy to accrete cold gas determines its continued ability to form stars and hence dictate where it falls amongst the Hubble sequence. Accreted gas, however, has many origins (such as filamentary `cold mode' accretion from the cosmic web, mergers or additionally cooling flows from a shocked hot halo) however is converted into stars within typical depletion timescales of order gigayears \citep{davis2016}. 
For material stripped in mergers or for shocked gas accreting through cooling flows, a natural consideration is that the accretion may not be necessarily aligned with the angular momentum of the benefiting galaxy \citep[e.g.][]{davis2011, lagos2015}. Misalignment can be considered to be a transient property as torques from the stellar component continually act to realign the gas component which can only be opposed by sustained misaligned accretion \citep[][]{vdvoort2015, davis2016}. 

Understanding the origin and nature of galaxies with decoupled star-gas rotation (kinematic misalignment - used interchangeably) has been the focus for several recent works. \citet{davis2011} found that $\sim 36$\% of ETGs exhibit misalignment between their star and gas rotation (i.e. difference of 30$^{\circ}$ between rotational axes) whose fraction increases for field ETGs. For late types, \citet{chen2016} first investigated star forming galaxies with counter-rotating stars and gas, a far rarer occurrence ($\sim$2\%), finding a clear boost in star formation in central regions. This suggests that the processes leading to misalignment are also inherent in cancelling angular momentum, enabling increased gas in-flows to nuclear regions. \citet{jin2016} extended this discussion to find that for a sample of 66 misaligned galaxies the misalignment fraction ($> 30^{\circ}$) is dependent on properties such as specific star formation rate, stellar mass and local environment, again determining that misaligned galaxies are more isolated. \citet{duckworth2019} explored the connection of misalignment in central galaxies to large scale environment (i.e. distance to cosmic web) and halo assembly time, finding that `cold' mode accretion from filaments was unlikely to contribute significantly to misalignment, noting that morphology held a far stronger relationship. \citet{bryant2019} considered misalignments for $\sim$1200 galaxies in the SAMI survey also demonstrating that, rather than local environment, that morphology held the strongest correlation with likelihood of star-gas decoupling. In simulations, \citet{starkenburg+19} considered the nature of counter-rotation in low mass galaxies in Illustris. They identify the key role of gas loss through black hole (BH) feedback and flyby interactions to enable misalignment through re-accretion of misaligned material. The mechanisms for decoupling gas are not fully determined and are likely a combination of both external and internal processes, both of which are seen to also shape the stellar angular momentum content of galaxies at $z=0$. To understand how these non-linear processes relate both to angular momentum retention from the dark matter halo and how this propagates to star-gas decoupling, a combination of both observations and simulations are required. 

This article is the first in a series which will comprehensively categorize the nature of galaxies that have decoupled rotation between their stars and gas. Utilizing a combination of both state of art IFS observations (MaNGA) and simulations (IllustrisTNG100) the aim of the project is to demonstrate the fundamental relationships behind this decoupling. Here we introduce our observational sample\footnote{Full catalogue of kinematic classifications will be made publically available after the final MaNGA data release (2021). Classifications for currently public MaNGA data can be made available on request. See \url{www.chrisduckworth.com} for catalogue and \S\ref{sec:obs_data} for description.} and mock sample in IllustrisTNG100. In this work we conduct a census of star-gas decoupling in both observations and simulations to study its link to galaxy morphology, stellar angular momentum and spin of its parent dark matter halo. In future companion papers we will explore the relationship of this kinematic decoupling with AGN and mergers.

This paper is structured as follows. Section \ref{sec:obs_data} describes the observational data we use in this work and our kinematic classifications. Section \ref{sec:sim_data} describes the simulation data and our construction of the mock sample. Section \ref{sec:manga_results} (Section \ref{sec:tng_results}) describes our results in MaNGA (IllustrisTNG100). Finally we discuss our findings in Section \ref{sec:discussion}, before concluding in Section \ref{sec:conclusion}.

\section{Observational Data} \label{sec:obs_data}
\subsection{The MaNGA survey}
Set to complete in 2020, the MaNGA survey is designed to investigate the internal structure of $\sim$10000 galaxies in the nearby Universe. By design, the complete sample is unbiased towards morphology, inclination and colour and provides a near flat distribution in stellar mass. 

MaNGA is one of three programs in the fourth generation of the Sloan Digital Sky Survey (SDSS-IV) which enables detailed kinematics through integral field unit (IFU) spectroscopy. MaNGA uses the SDSS 2.5 metre telescope in spectroscopic mode \citep{gunn2006} with the two dual-channel BOSS spectrographs \citep{smee2013} and the MaNGA IFUs \citep{drory2015}. MaNGA provides spatial resolution on kpc scales (2'' diameter fibres) while covering 3600-10300$\mathring{A}$ in wavelength with a resolving power that varies from R$\sim$1400 at 4000$\mathring{A}$ to R$\sim$2600 at 9000$\mathring{A}$. 

MaNGA observations are covered plate by plate, employing a dithered pattern for each galaxy corresponding to one of the 17 fibre-bundles of 5 distinct sizes. Any incomplete data release of MaNGA should therefore be unbiased with respect to IFU sizes and hence a reasonable representation of the final sample scheduled to be complete in 2020.

\begin{figure}
	\includegraphics[width=\linewidth]{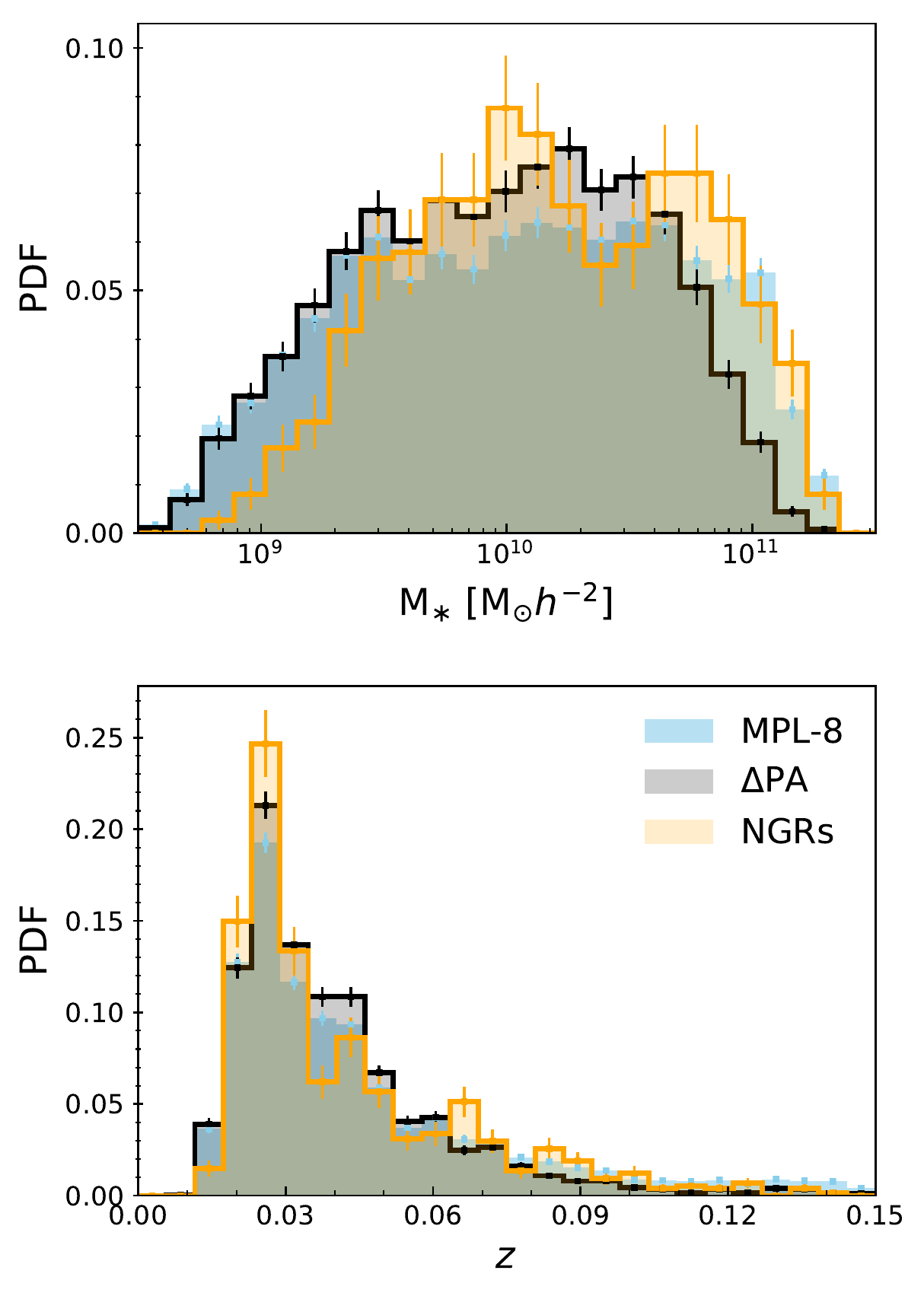}
    \caption{Relative frequency distributions of stellar mass and redshift for the MaNGA MPL-8 sample (light blue), our $\Delta$PA defined sample (black) and those with a defined stellar PA but no clear H$\alpha$ rotation (orange). The figure is cut at $z=0.15$ representing the extent of MaNGA targets. Each histogram is given with Poisson errors on each bin.}
    \label{fig:samp_cons}
\end{figure}

The majority of observations contribute to one of the three main subsets: the Primary sample, the Secondary sample and the Colour-Enhanced supplement. All sub-samples observe galaxies to a minimum of $\sim 1.5$ effective radii ($R_{e}$) with the Secondary sample increasing this minimum to $\sim 2.5 R_{e}$. The Colour-Enhanced supplement fills in gaps of the colour-magnitude diagram leading to an approximately flat distribution of stellar mass. A full description of the MaNGA observing strategy is given in \citet{law2015obs,yan2016obs}. 
The raw observations are processed by the MaNGA Data 
Reduction Pipeline (DRP) as described in \citet{law2016drp, yan2016spec}. The fibre flux and inverse variance is extracted from each exposure, which are then wavelength calibrated, flat-fielded and sky subtracted. In this work, we use 6044 galaxies from the eighth MaNGA Product Launch (MPL-8) that are selected in the Primary, Secondary and Colour-Enhanced samples and have non-critical observations. Figure \ref{fig:samp_cons} shows the distribution of stellar mass and redshift of the eighth MaNGA Product Launch (MPL-8) with comparison our $\Delta$PA defined sample and those with coherent stellar rotation but no clear H$\alpha$ rotation (NGRs) outlined in \S\ref{sec:visual_classifications}.

\subsection{Spectral fitting for kinematics}
All stellar and H$\alpha$ velocity fields are taken directly from the MaNGA Data Analysis Pipeline \citep[DAP;][for an overview and emission line modelling respectively]{westfall2019, belfoire2019}, we direct the reader to these references, however we summarise the key points here.

The DAP extracts stellar kinematics using the Penalised Pixel-Fitting (pPXF) method \citep{cappellari2004,cappellari2017}. The DAP fits the stellar continuum of each spaxel to extract the line of sight velocity dispersion and then fits the absorption-line spectra from a set of 49 clusters of stellar spectra from the MILES stellar library \citep{sanchez2006,falcon2011}. Before extraction of the mean stellar velocity, the spectra are spatially Voronoi binned to $g$-band \textit{S/N} $\sim$ 10, excluding any individual spectrum with a $g$-band \textit{S/N} < 1 \citep{cappellari2003}. This approach is geared towards stellar kinematics as the spatial binning is applied to the continuum \textit{S/N}, however, we note that unbinned and Voronoi binned velocity maps produce similar results. 

Ionized gas kinematics are extracted by the DAP through fitting a Gaussian to the H$\alpha$-6564 emission line, relative to the input redshift for the galaxy. This velocity is representative for all ionized gas, since the parameters for each Gaussian fit to each emission line are tied during the fitting process. These velocities are also binned spatially by the Voronoi bins of the stellar continuum. 

\subsection{Defining global position angles}
For a complete description of PA fitting and typical error estimation for MaNGA, we direct the reader to \citet{duckworth2019}. Here we use a similar process, summarising the key points and highlighting differences.

Global position angles (PA) are estimated for both the stellar and ionized gas velocity fields using the \texttt{fit\_kinematic\_pa} routine \citep[see Appendix C of][for a description of the process]{krajnovic2006}. \texttt{fit\_kinematic\_pa} returns the angle (counter-clockwise) of the bisecting line which has greatest velocity change along it. The best fit angle is found by sampling 181 equally spaced steps, so that the output PA will have precision of 0.5$^{\circ}$. By default, \texttt{fit\_kinematic\_pa} returns a PA defined between 0$^{\circ}$ and 180$^{\circ}$, which is indiscriminate towards direction of the blueshifted and redshifted sides. To adjust this, we identify the redshifted side and return PAs defined by the angle to the redshifted side clockwise from the north axis (0-360$^{\circ}$). 

The accuracy of PA fitting is biased by neighbouring galaxies, spectral pixels (spaxels) with spuriously high velocities and inclination. 

Foreground stars are removed during the spectral fitting, however foreground/background galaxies can remain within the IFU footprint. This can be a significant problem for global PA fitting since \texttt{fit\_kinematic\_pa} symmetrizes the velocity fields and interpolates to estimate the PA. Background/small objects can then bias the PA fit for the main target, especially in the instance where they are moving significantly different to the target galaxy (e.g. when they are at different $z$). To counteract this, we remove all disconnected regions smaller than $10\%$ of the target galaxy's footprint. 

Spaxels with spuriously high velocities (e.g. > 1000km/s relative to target's central redshift) can also bias PA fits during symmetrization. These often correspond to background galaxies that are connected (on the sky) to the target galaxy's footprint, and hence, we sigma clip the velocity field and remove all spaxels above a $3\sigma$ threshold.

Accurate PA estimation is naturally more difficult for near edge-on galaxies. disc obscuration and a smaller surface area allow individual Voronoi bins to more easily bias overall PA fits. This inherently leads to a larger scatter in PA fitting around the true value, especially due to central spaxels during symmetrization.

\subsection{Visual Classifications} \label{sec:visual_classifications}
Global position angles are only well defined for coherently rotating velocity fields. Those with a decoupling between inner and outer regions due to warps or kinematically decoupled cores (KDCs) are poorly described by global PAs. 

To select a clean sample of galaxies with well defined global PAs, we visually classify all of the velocity fields after pre-processing and PA fitting. Both stellar and H$\alpha$ velocity fields are characterised into 3 categories;
\begin{itemize}
    \item Dominant coherent rotation and well defined PA.
    \item Dominant coherent rotation but with higher noise or more complex motion resulting in a usable PA fit but with higher typical errors. Highly inclined velocity fields with a higher likelihood of biased PAs fits are included in this category. 
    \item Do not use.
\end{itemize}

Kinematic features are also identified. Both stellar and H$\alpha$ velocity fields are classified into;
\begin{itemize}
    \item Kinematically decoupled core (i.e. those with a central component that rotates in a different direction ($> 30^{\circ}$) with respect to the overall galaxy rotation)
    \item Warp (velocity field of central region is warped with respect to outskirts)
    \item Merger (ongoing merger or neighbour identified within IFU. Only those with obvious disruption are followed up in photometry.)
    \item No feature
\end{itemize}
The majority of those with kinematic features have poorly defined global PAs and hence are flagged as do not use for the previous flag. 

\begin{figure*}
	\includegraphics[width=0.87\linewidth]{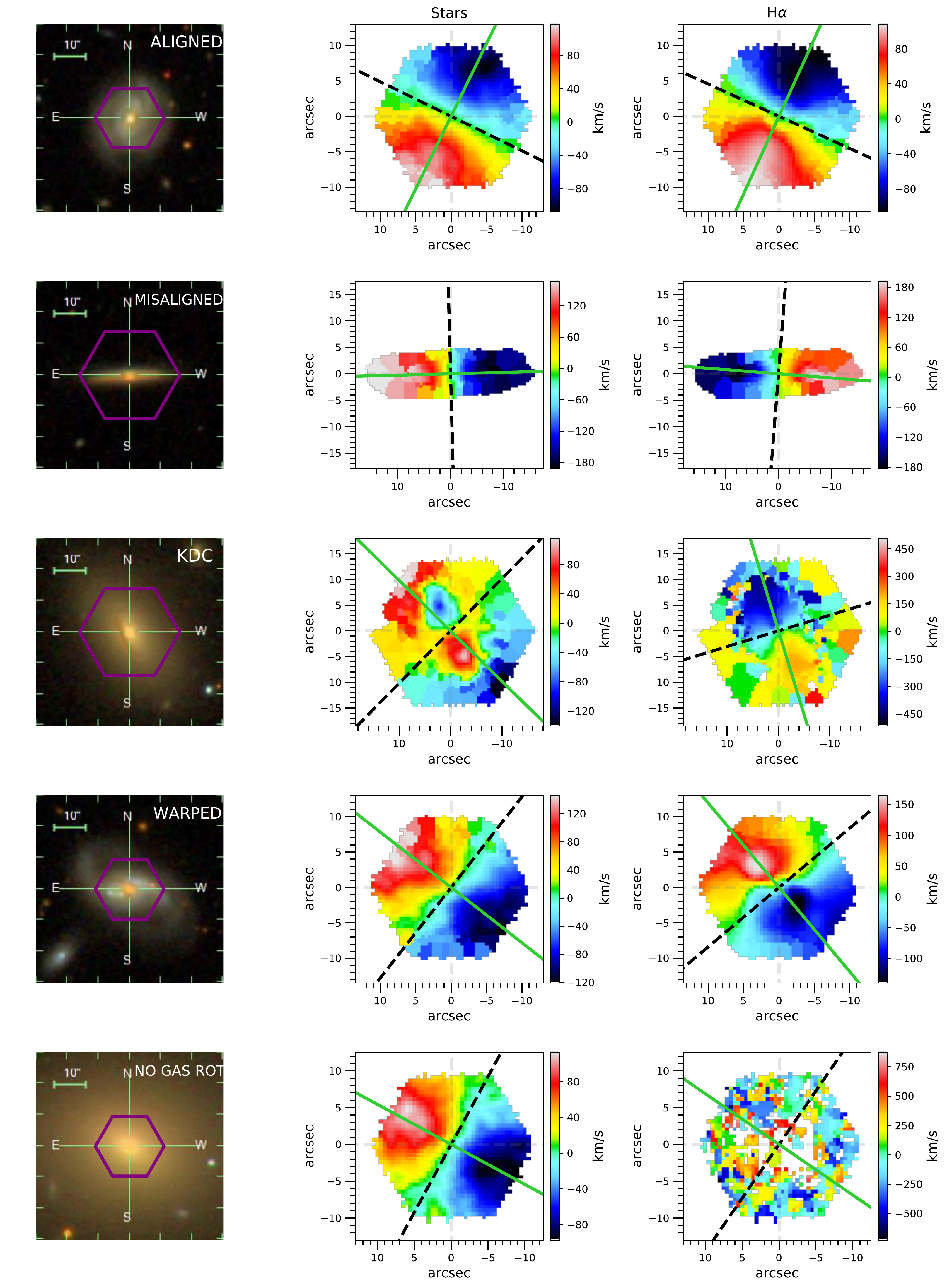}
    \caption{Examples of PA fits for galaxies with different kinematic classifications. For each galaxy (row), we show the photometry taken from SDSS with the MaNGA IFU observation footprint overlaid in purple (left), the stellar velocity field (middle) and the H$\alpha$ velocity field (right). The kinematic PA fits (see \S\ref{sec:def_kin_mis}) are shown on the velocity fields (green solid line) with the axis of rotation (black dotted line). The kinematic classifications from top to bottom are; (a) PLATEIFU: 7958-6101, kinematically aligned near face on; (b) PLATEIFU: 8465-12704, counter-rotating near edge on; (c) PLATEIFU: 9868-12704, with a KDC in the stellar velocity; (d) PLATEIFU: 8252-6103, with a warped H$\alpha$ velocity field with respect to the stellar; (e) PLATEIFU: 10219-6102, with a centrally depleted/missing H$\alpha$ velocity field but coherent rotation in the stellar.}
    \label{fig:mis_grid}
\end{figure*}

For studies of quenching it may be useful to consider galaxies that have defined stellar rotation but lack coherent motion in the ionized gas. For galaxies that have usable PAs for the stellar velocity but unusable PAs for the ionized gas, we define a further classification of the gas velocity field;
\begin{itemize}
    \item Depletion (seen as empty spaxels signifying lack of gas, usually in central regions)
    \item No clear rotation (map has no clear rotation or is noise dominated)
    \item Biased rotation (partial rotation in the velocity field, however there are significant regions with incoherent motion)
    \item No clear characteristics/ No gas.
\end{itemize}
We note there is a clear overlap between the classifications for depletion and no clear rotation, since velocity fields are often a combination of these two features. The total numbers for each classification in each category are summarised in Table \ref{tab:kin_class}. Examples of PA fits (see \S\ref{sec:def_kin_mis} for calculation) with the associated photometry for various kinematic classifications is given in Figure \ref{fig:mis_grid}. Examples of galaxies that are kinematically aligned, misaligned, have a stellar kinematically decoupled core, have a warped H$\alpha$ velocity field and have clear stellar rotation but depleted ionized gas/ no rotation are shown respectively. 

\begin{table*}
\begin{tabular}{lrrrrrrlll}
\hline
&  Clean PA &  Messy PA &  Unusable PA &  KDC &  Warp &  Merger & Depletion & No clear rotation & Biased rotation \\
\hline
Stellar &      3290 &      1581 &         1172 &   47 &    39 &     116 &       960 &               960 &             960 \\
H$\alpha$ &      2876 &      1071 &         2097 &   17 &    82 &     116 &       562 &               180 &             175 \\
Both &      2848 &      1023 &         1136 &    5 &    11 &     116 &        -- &                -- &              -- \\ 
\end{tabular}
\caption{Summary table of galaxy numbers for kinematic classifications in MPL-8. Each row shows the total number of galaxies in the classification criteria defined in each column for stellar velocity fields only, gas velocity fields only and both (top to bottom). Columns 1-3 correspond to the quality of the PA fit, 4-6 correspond to kinematic features and 7-9 correspond to additional notes for the H$\alpha$ velocity field (see text for details about classifications). Columns 7-9 are only defined for unusable PAs for H$\alpha$ and clean/messy PAs for the stellar field. The total number of galaxies meeting this criteria is given in the stellar row for columns 7-9.}
\label{tab:kin_class}
\end{table*}

\subsection{Defining kinematic misalignment} \label{sec:def_kin_mis}
Only selecting galaxies with dominant coherent rotation (both clean and messy) for both stellar and H$\alpha$ velocity fields with no defined features in either map, we are left with 3798 galaxies used in this analysis. The mass distribution of the $\Delta$PA defined sample with respect to MPL-8 is shown in Figure \ref{fig:samp_cons}. We define the offset angle between the kinematic PAs of the stellar and ionized gas fields as such; 
\begin{equation} \label{eq:delPA}
\Delta PA = |PA_{stellar} - PA_{H\alpha}|. 
\end{equation}
We define galaxies with $\Delta$PA $\geq 30^{\circ}$ to be significantly kinematically misaligned. The choice while somewhat arbitrary, is chosen for comparison with previous literature \citep[e.g.][]{davis2011, bryant2019}. Regardless 30$^{\circ}$ represents a conservative choice for selecting galaxies with significant decoupling; the reasons for which are twofold. Firstly, since we are comparing the ionized gas in MaNGA to all gas in IllustrisTNG100, we must take into the account different kinematic properties of different gas phases. In observation, \citet{davis2011molecular} find that the typical difference between the PAs of cold gas (CO) and ionized gas can be described by a Gaussian distribution centred on 0 with a standard deviation of 15$^{\circ}$ for 38 CO bright galaxies in ATLAS\textsuperscript{3D}. While indicating ionized gas is a reasonable estimator for cold gas, splitting at $\Delta$PA = 30$^{\circ}$ accounts for the scatter in this relationship. Secondly the split takes into account both the error in $\Delta$PA estimation \citep[a few $^{\circ}$; see Appendix A3 of][]{duckworth2019} and projection effects since it is a projection of a 3D property. We note that taking a different split at 40$^{\circ}$ does not change any of our findings.

\subsection{Morphology} \label{sec:morph_def}
We classify the morphology of MaNGA galaxies through the formalism laid out by the citizen science project; GalaxyZoo2 \citep[GZ2;][]{willett2013}. GZ2 provides visually identified morphologies (and also measures finer morphological features e.g. bars, bulge size and edge-on discs) for 304,122 galaxies drawn from SDSS. GZ2, however, is not complete for the MaNGA sample and has been combined with an unpublished version; GalaxyZoo4 with debiasing code re-run to provide a consistent set of definitions for all MaNGA targets (see \url{https://www.sdss.org/dr15/data_access/value-added-catalogs/?vac_id=manga-morphologies-from-galaxy-zoo}). 

In a nutshell, GZ2 provides morphological classification through a decision tree of questions. Further questions are dependent on the answer to the previous to characterise a certain morphological type and identify finer features (see Figure 1 in \citet{willett2013} for this flowchart). From this, a table of vote fractions for each question combined with the total number of votes dictate a reliably sampled galaxy population with a set of desired morphological features. 

The first question in the decision tree 'Is the galaxy smooth and rounded with no sign of a disc?', allows categorisation into broad ETGs and LTGs. We select galaxies with a debiased vote fraction > 0.7 for smooth to be ETGs and galaxies with a debiased vote fraction of > 0.7 for disc or features to be LTGs. Defining an exact population of lenticular galaxies (S0s) is tricky through public classifications. LTGs, however, can be separated based on the dominance of the bulge with respect to the disc in GZ2 through the question 'How prominent is the central bulge, compared with the rest of the galaxy?'. \citet{willett2013} demonstrate a strong correlation between bulge dominance as defined per this question and expert classifications of T-type \citep{nair2010}. Equation 19 of \citet{willett2013} provides a linear mapping from GZ2 bulge classification to expert defined morphological T-type. Care must be taken in using this linear mapping \citep[see discussion in][]{willett2013}, however, this should be a reasonable parameterisation to coarsly separate LTGs into earlier-type (S0 - Sa) and later-type spirals (Sb - Sd). We split our LTG population at T-type = 3, to give three morphological categories along with pure ETGs. 

The estimates of gas mass used here for MaNGA are derived from the Pipe3D pipeline \citep{pipe3Da, pipe3Dvac}, which uses dust attenuation within the footprint of the IFU, which methodology is described in \citet{barrera2018}.

\subsection{Group membership} \label{sec:group_def}
To investigate different pathways leading to kinematic misalignment, we must separate galaxies into centrals and satellites. We identify groups with an adaptive halo-based group finder of \citet{yang2005,yang2007} and with improved halo mass assigning techniques \citep[see;][for details and application to SDSS]{lim2017}. In a nutshell, the group finder uses either the stellar mass or luminosity of central galaxies in addition with the $n^{th}$ brightest/most massive satellite as proxies for halo mass. Galaxies are assigned to groups through an iterative process, where halo properties such as halo size and velocity dispersion are updated until membership converges. 



\citet{lim2017} do not apply the group finder to the thin strips in the Southern Galatic Cap of SDSS main due to incomplete groups resulting from close proximity to borders. MaNGA galaxies in these strips are therefore unclassified by the group finder, resulting in 5088 matched galaxies with group membership classifications into central or satellite. 

\section{Simulation data} \label{sec:sim_data}
\subsection{IllustrisTNG}
The IllustrisTNG project \citep{marinacci18,naiman18,nelson18,pillepich18b,springel18} is a suite of magneto-hydrodynamic cosmological scale simulations incorporating an updated comprehensive model for galaxy formation physics \citep[as decribed in][]{weinberger17,pillepich18a} and making use of the moving-mesh code \texttt{AREPO} \citep{springel10,pakmor11,pakmor13}. For this work, we use the highest resolution fiducial run of TNG100 which follows the evolution of 2 x 1820$^3$ resolution elements within a periodic cube with box lengths of 110.7 Mpc (75 h$^{-1}$ Mpc). This corresponds to an average mass resolution of baryonic elements of 1.4 x 10$^6 M_{\odot}$ and 7.5 x 10$^6 M_{\odot}$ for dark matter. Here we make use of public data from the IllustrisTNG project \citep[as described in][]{nelson2019}.

Structure in TNG100 is identified into haloes and subhaloes as follows. Haloes (also referred to as FoF haloes or Groups) are found from a standard friends-of-friends (FoF) algorithm \citep{davis85} with linking length $b=0.2$. The FoF algorithm is run on the dark matter particles, and the other types (gas, stars, BHs) are attached to the same groups as their nearest DM particle. Each halo is then divided into gravitationally bound subhaloes through the subfind algorithm \citep{springel01}. In short, subfind defines `subhaloes' as locally over-dense and self-bound particle groups as distinct objects within given FoF haloes. We consider all subhaloes at $z=0$ containing a minimum stellar mass of $M_{\ast} = 10^{8.5} M_{\odot}$ to potentially make up our mock MaNGA like sample. Since we are typically considering the stellar component of these subhaloes for our mock observations, we will refer to these as TNG100 galaxies.

\subsection{Matching to MaNGA sample}
To construct a mock MaNGA sample we select representative subhaloes from TNG100. For every MaNGA galaxy, we find the TNG100 galaxy with the most similar stellar mass, size and SDSS $g - r$ colour. In this instance, stellar mass is defined by the total mass of stellar particles within a radius of 2 stellar effective radii. The SDSS $g - r$ colour is found using the prescription outlined in \citet{nelson18}. Here we describe the general process, while we direct the reader to \citet{nelson18} for more detail. Each stellar particle in the simulation is modelled as a single-burst simple stellar population. This is converted into a population spectrum using FSPS \citep{conroy2009,conroy2010,foreman_mackey2014} which is convolved with the pass-bands for SDSS colours. We use model C \citep[as described in][]{nelson18} which also includes models for unresolved and resolved dust. We use sizes following the prescription of \citet{genel2018}, which use a projected half light radius. The SDSS bands are constructed as above and are used to define circular half light radii for each SDSS band along X, Y and Z projections of the box. We use the $r-$band half light radius projected perpendicular to the XY plane, consistent with the line of sight of the mock MaNGA observation.

The matching is done through finding the closest neighbour in a normalised space with dimensions of the matched properties. If multiple MaNGA objects match to a given TNG100 galaxy then the absolute nearest neighbour is selected and the MaNGA object is assigned to its second nearest neighbour. The process is iterated until all have unique matches. 

The galaxy is then assigned the same bundle size IFU as the matched MaNGA galaxy with the corresponding angular resolution. The galaxy is then `observed' (see \S\ref{sec:mock_obs}) at a distance so that the angular footprint of the assigned IFU covers the same number of physical effective radii for the mock galaxy as the matched observation. 

\subsection{Mock observations} \label{sec:mock_obs}
We convert each galaxy in TNG100 into a mock MaNGA observation, as follows:

We take the raw particle/cell data of gas and stars and project on the XY plane (i.e. z-direction is the line of sight). Since there is no preferred direction in the simulation, this corresponds to a 'random' viewing angle of each galaxy. We bin particles corresponding to the angular resolution of spaxels in MaNGA (0.5 arcsec/pixel), in the distinct hexagonal fibre bundle footprint. In each bin, we calculate the mean velocity, velocity dispersion and total flux for all particles. 

Since we include all particles along the line of sight, we must take care in interpreting the absolute values of flux, since none is lost due to obscuration. We, however, do not use the flux values calculated here in our work.

\begin{figure}
	\includegraphics[width=\linewidth]{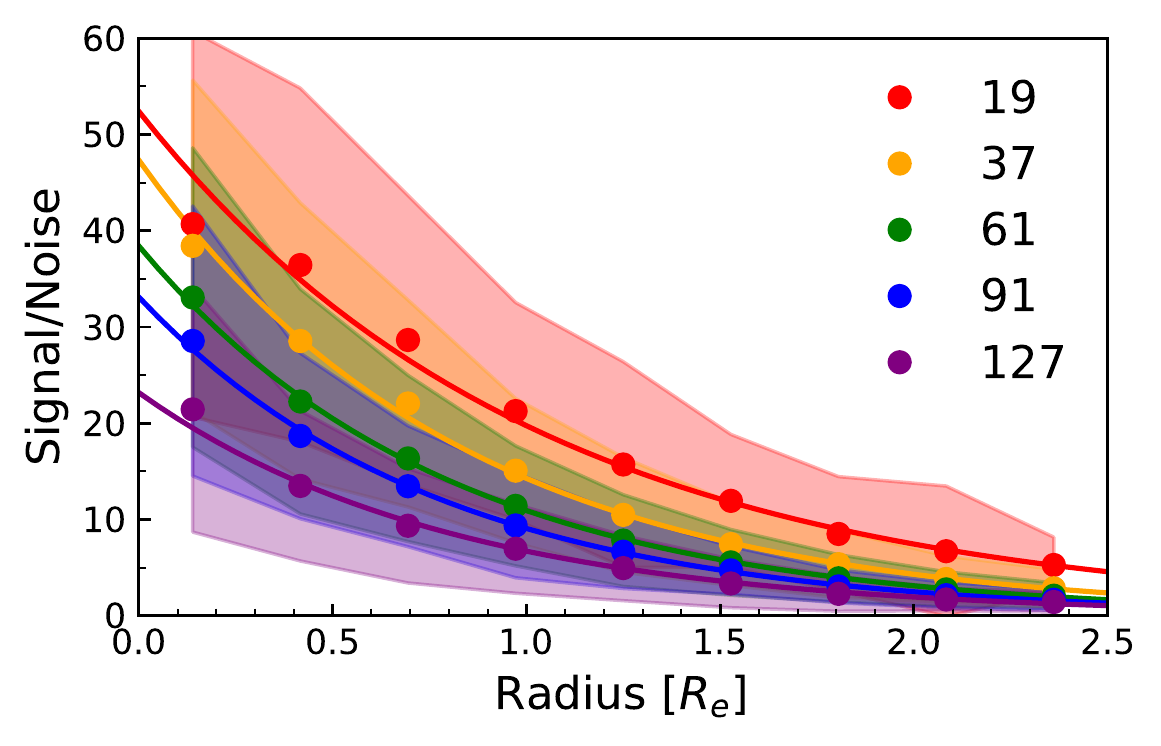}
    \caption{Average signal to noise profiles for each IFU size for all MaNGA MPL-8 observations. The circles show the median value for each radius bin with the shaded region corresponding to the standard deviation. The solid line corresponds to an logarithmic parametric fit to the data points, used in sampling the noise profile for the mock observations.}
    \label{fig:noise_profile}
\end{figure}

In order to estimate the typical noise associated with a MaNGA observation, we compute radial profiles of the signal to noise ratio (SNR) for all MaNGA observations of a given IFU fibre bundle size. MaNGA provides estimates of the SNR for every spaxel in each observation in the $g$-band. Figure \ref{fig:noise_profile} shows the azimuthally averaged SNR profiles for all MaNGA observations of each fibre bundle size. We fit a logarithmic function to each profile, which is used to assign noise to the mock observations. Noise is drawn for each pixel from a normal distribution using the median and standard deviation of the fitted logarithmic radial profile.   

In order to simulate the effects of the point spread function (PSF), we then convolve our binned particle data with a Gaussian kernel. MaNGA observations typically have a $g$-band PSF which can be fit with a Gaussian of $\sim 2-3''$ full width half maximum (FWHM). We take all our mock observations to have a PSF modelled by a Gaussian with a 2$''$ FWHM. 

\begin{figure*}
	\includegraphics[width=\linewidth]{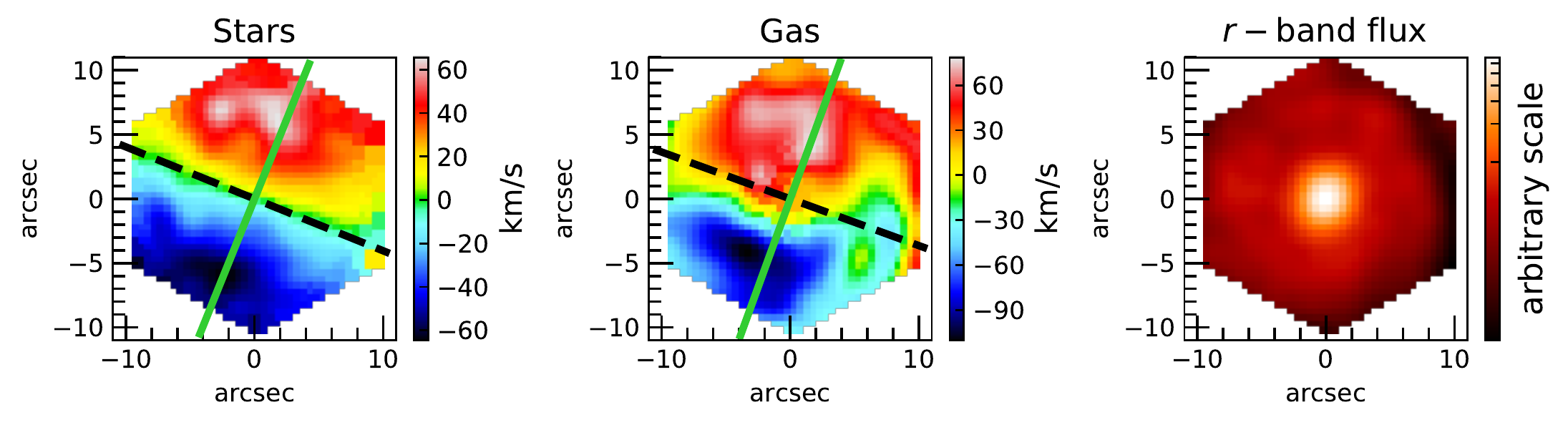}
    \caption{Example outputs from a MaNGA-like observation in TNG100. Shown (left to right) are the stellar velocity field, gas velocity field and normalised $r-$band flux for a given galaxy, `observed' under the same conditions of its MaNGA counterpart (i.e. distance and IFU size). For the stellar and gas velocity fields, the kinematic PA fits (see \S\ref{sec:def_kin_mis}) are shown (green solid line) with the axis of rotation (black dotted line).}
    \label{fig:example_obs}
\end{figure*}

We fit position angles to MaNGA observations that have been Voronoi binned so that bins contain a minimum S/N $\sim 10$. To maintain consistency and avoid spurious individual particles biasing measurements, we also Voronoi bin our mock observations so that a minimum of 5 particles is contained within a given bin, again using the routine of \citet{cappellari2003}. Figure \ref{fig:example_obs} shows example stellar (and gas) velocity and dispersion fields along with normalised $r$-band flux, after our processing. 

\section{Misaligned galaxies in MANGA} \label{sec:manga_results}
\subsection{Total population} \label{sec:manga_total_pop}
\begin{figure}
	\includegraphics[width=\linewidth]{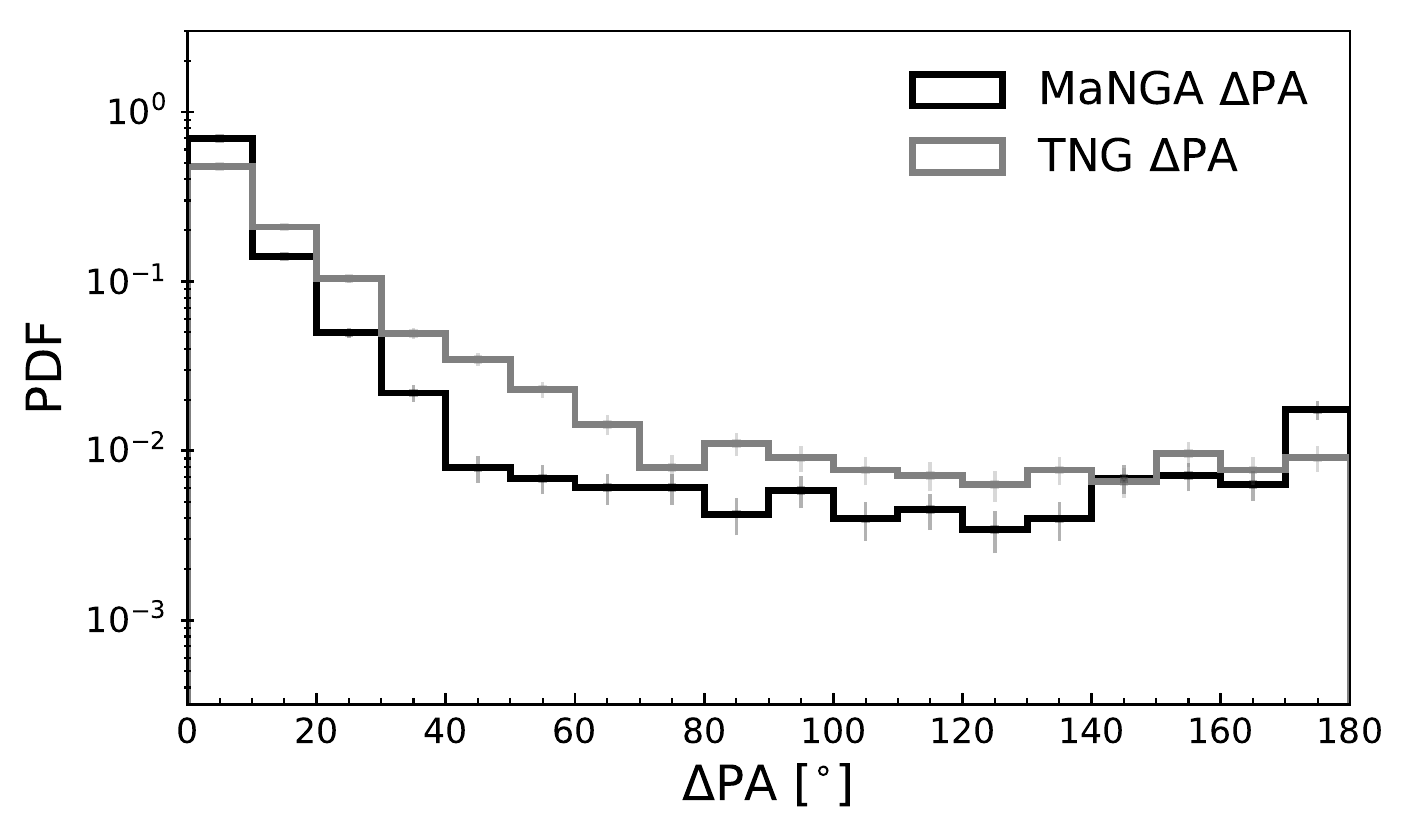}
    \caption{Probability density distribution of kinematic misalignment as defined by $\Delta$PA for the total MaNGA sample (black line) and matched TNG100 sample (grey line). $\Delta$PA is strongly peaked around 0$^{\circ}$ with a small boost close to 180$^{\circ}$.}
    \label{fig:total_pa_dist}
\end{figure}

Firstly we consider all $\Delta$PA defined galaxies for both MaNGA and TNG100. Figure \ref{fig:total_pa_dist} shows the distribution of $\Delta$PA for both MaNGA and IllustrisTNG100. Both distributions are strongly peaked around around 0$^{\circ}$ indicative of the preferentially aligned state predicted from TTT. The MaNGA distribution shows a sharp drop-off past 40$^{\circ}$ whereas TNG100 shows a smoother drop off to higher misalignments. Additionally the MaNGA distribution shows a second peak around 180$^{\circ}$ indicative of the stable counter-rotating state identified in previous work \citep[e.g.][]{chen2016}. This secondary peak is not seen for the overall TNG100 sample, however is apparent for star forming galaxies in TNG100 (see bottom panel of Figure \ref{fig:group_morph_PA}). 

\begin{figure}
	\includegraphics[width=\linewidth]{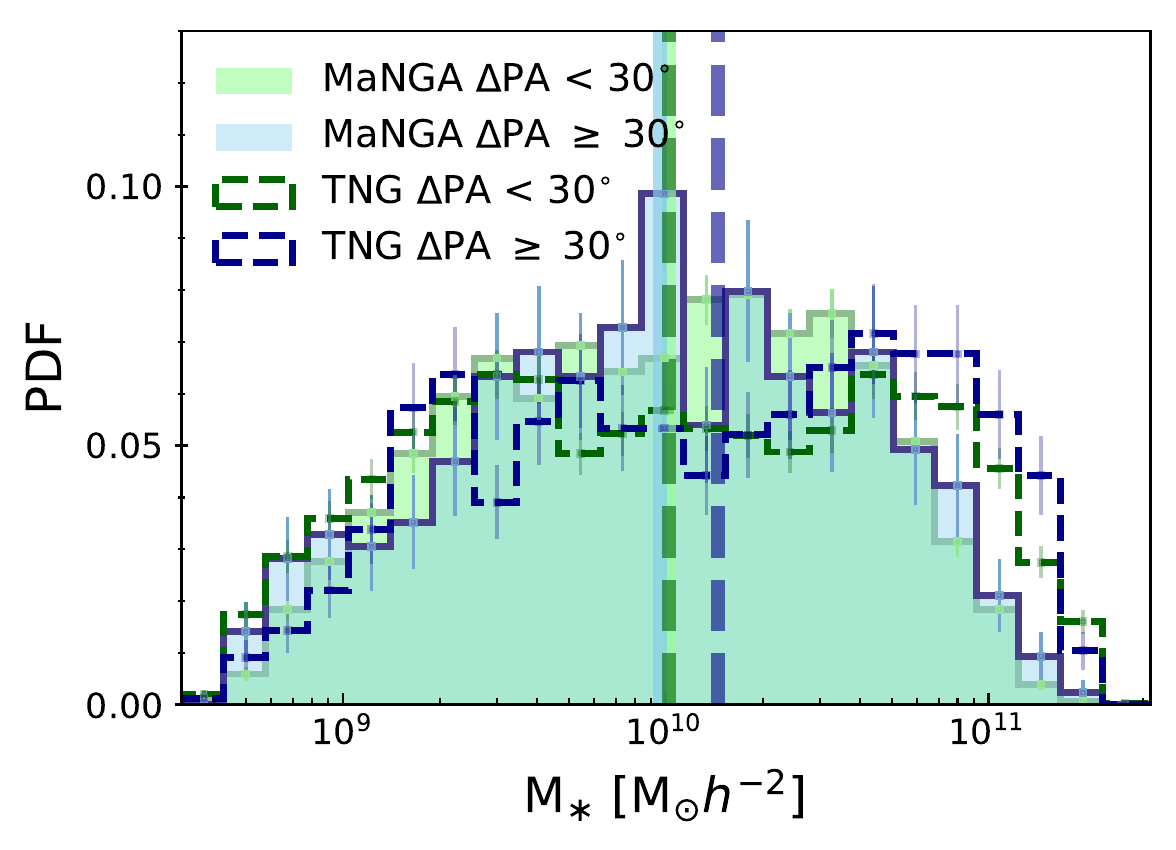}
    \caption{Probability density distributions of stellar mass, $(M_{\ast}/M_{\odot})$ for aligned galaxies ($\Delta$PA < 30$^{\circ}$, green) and misaligned galaxies ($\Delta$PA > 30$^{\circ}$, red) defined in MPL-8 (solid lines) and TNG100 (dashed). The vertical lines denote the corresponding distribution's median. The overall distributions are a reasonable match between mocks and observations, with a noted preference for $\Delta$PA defined galaxies at the very high mass end for TNG100. }
    \label{fig:TNG_mpl8_stelM}
\end{figure}

The TNG100 mock sample reproduces the general trends well, when considering the differences in how we split the samples in observations and simulations. The smoother drop-off past 40$^{\circ}$ for TNG100 is likely a combination of how we construct the mock observations and scatter in the mass distributions between the MaNGA and TNG100 samples. By construction the matching between MaNGA and TNG100 objects is done before $\Delta$PA is calculated. For this reason there may be differences between the mass distribution of the $\Delta$PA defined MaNGA and TNG100 samples, as shown in Figure \ref{fig:TNG_mpl8_stelM}. We find that the misaligned sample in TNG100 is slightly more massive with respect to MaNGA whereas the aligned samples are consistent. Due to the strong morphological dependence on kinematic misalignment, there is a secondary dependence on stellar mass. The increased overall fraction of misaligned galaxies in TNG100 is therefore, in part, due to the TNG100 $\Delta$PA defined sample being slightly more massive.

For the rest of this section, we will only consider the properties of MaNGA galaxies leaving the results of the TNG100 mock sample to \S\ref{sec:tng_results}.

\begin{figure}
	\includegraphics[width=\linewidth]{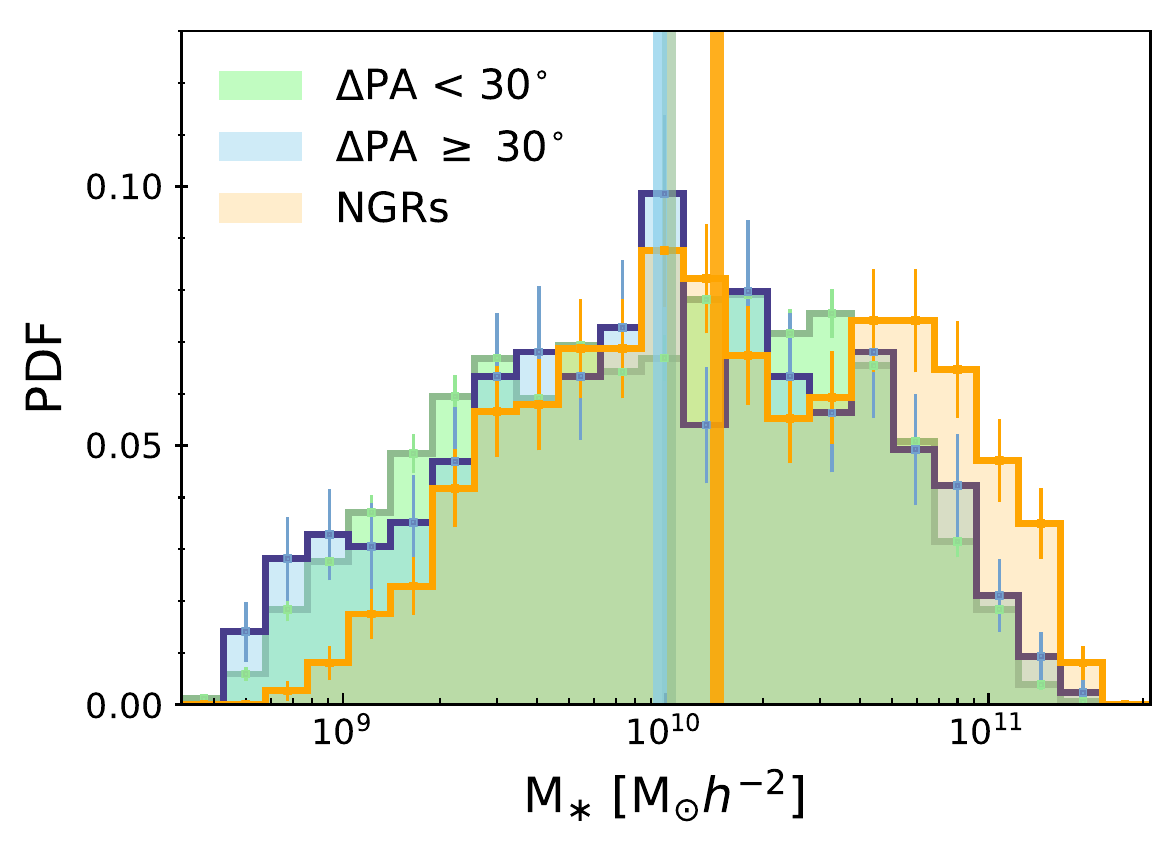}
    \caption{Probability density distributions of stellar mass, $(M_{\ast}/M_{\odot})$ for aligned galaxies ($\Delta$PA < 30$^{\circ}$) shown in green, those with high misalignment ($\Delta$PA > 30$^{\circ}$) are in blue and NGRs are in orange. Each histogram is given with Poisson errors on each bin. The vertical lines denote the corresponding distribution's median. NGRs are typically at higher stellar mass than those with aligned star and gas rotation.}
    \label{fig:delPA_stelM}
\end{figure}

We divide our MaNGA $\Delta$PA defined population at $\Delta$PA = 30$^{\circ}$ into aligned and misaligned. In the following, we also consider galaxies with defined stellar PAs but undefined H$\alpha$ due to central depletion or incoherent rotation/dispersion domination (no gas rotation; NGRs). Figure \ref{fig:delPA_stelM} shows the distribution of stellar mass for these three populations. We see no significant difference between the aligned and misaligned galaxies, however NGRs appear to be slightly more massive. \citet{graham2018} previously demonstrated the tight correlation between stellar angular momentum and stellar mass for MaNGA (MPL-5; $\sim$2300 galaxies). Since NGRs and misaligned galaxies are slightly higher mass, it could be expected that they are typically less rotationally supported with respect to the $\Delta$PA defined populations. 

Here we use the luminosity weighted stellar angular momentum estimator, $\lambda_R$, taken directly from Equation 1 in \citet{emsellem2007} as
\begin{equation}
\lambda_{R} \equiv \frac{\langle R | V | \rangle}{ \langle R \sqrt{ V^{2} + \sigma^{2} } \rangle } = \frac{ \Sigma_{ n = 1 }^{ N } F_{ n } R_ { n } \left| V_{ n } \right| }{ \sum_{ n = 1 }^{ N } F_{n} R_{ n } \sqrt{ V_{ n }^{ 2 } + \sigma_{ n }^{ 2 } } }.
\end{equation}
$\lambda_R$ is calculated from summing over N pixels in the IFU observation within the radius of interest $R$. $F_{n}$, $V_{n}$ and $\sigma_{n}$ are the flux, line of sight velocity and line of sight velocity dispersion of the nth pixel. Here we present all measures of $\lambda_R$ encompassing a radius of $1.5R_e$ weighted by $r-$band flux. We also take the ellipticity to be $\epsilon = 1 - b/a$ where $a$ and $b$ are the major and minor axes of the galaxy estimated from the NASA Sloan Atlas catalogue \cite[used for target selection in MaNGA;][]{blanton2011}.

\begin{figure*}
	\includegraphics[width=\linewidth]{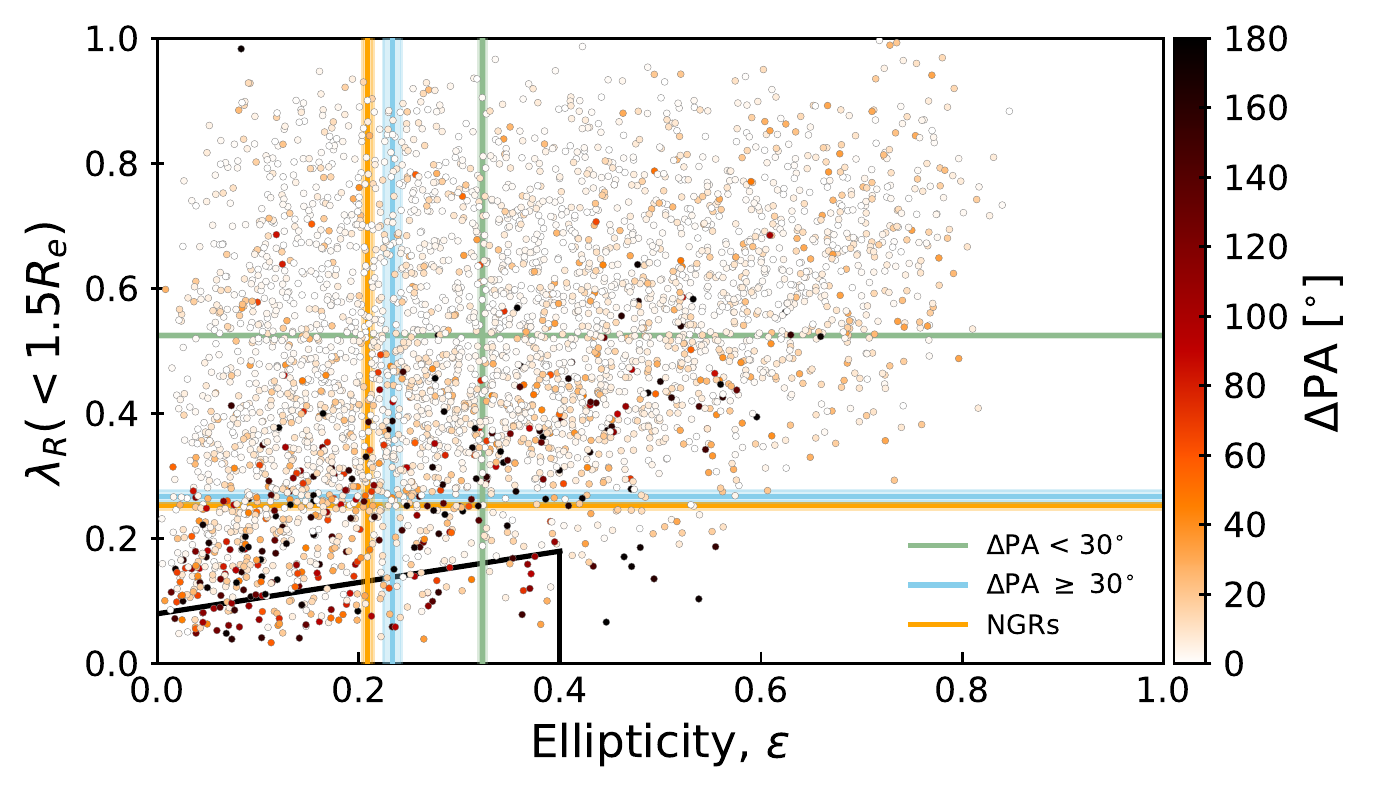}
    \caption{$\lambda_R$ within 1.5$R_e$ against ellipticity, $\epsilon$ for all galaxies with defined $\Delta$PA. The individual points are for all $\Delta$PA defined MaNGA galaxies coloured by $\Delta$PA according to the colorbar. Medians for kinematically aligned ($\Delta$PA < 30$^{\circ}$), misaligned ($\Delta$PA > 30$^{\circ}$) and NGRs are shown by the green, light blue and orange lines respectively. The lighter shade around each line corresponds to the standard error. Aligned galaxies reside more typically in the fast rotator regime with higher $\lambda_R$ and $\epsilon$, whereas misaligned galaxies and NGRs reside closer to the slow rotator regime. The same qualitative trends are found if this plot is made for ETGs, S0-Sas or Sb-Sds alone.}
    \label{fig:delPA_lambda_Re}
\end{figure*}

Figure \ref{fig:delPA_lambda_Re}, shows $\lambda_R$ vs $\epsilon$ for all $\Delta$PA defined galaxies and the medians for the aligned, misaligned and NGR samples. Kinematically aligned galaxies reside at preferentially higher $\lambda_R$ and ellipticity with respect to NGRs. This is indicative of the dispersion dominance over rotation for disrupted gas poor and typically higher mass galaxies that we see in our NGR sample. Interestingly, kinematically misaligned galaxies also typically reside close to the slow rotator regime (defined by the black solid line). In addition, the same qualitative trends are seen (i.e. misaligned and NGR galaxies have lowered angular momentum with respect to the aligned) are seen if this plot is made for ETGs, S0-Sas or Sb-Sds alone. The fast/slow rotator classification refers to whether a given galaxy's rotation can be considered regular (circular velocity) or exhibits dispersion dominated motion \citep[][]{emsellem2007}.

\begin{figure}
	\includegraphics[width=\linewidth]{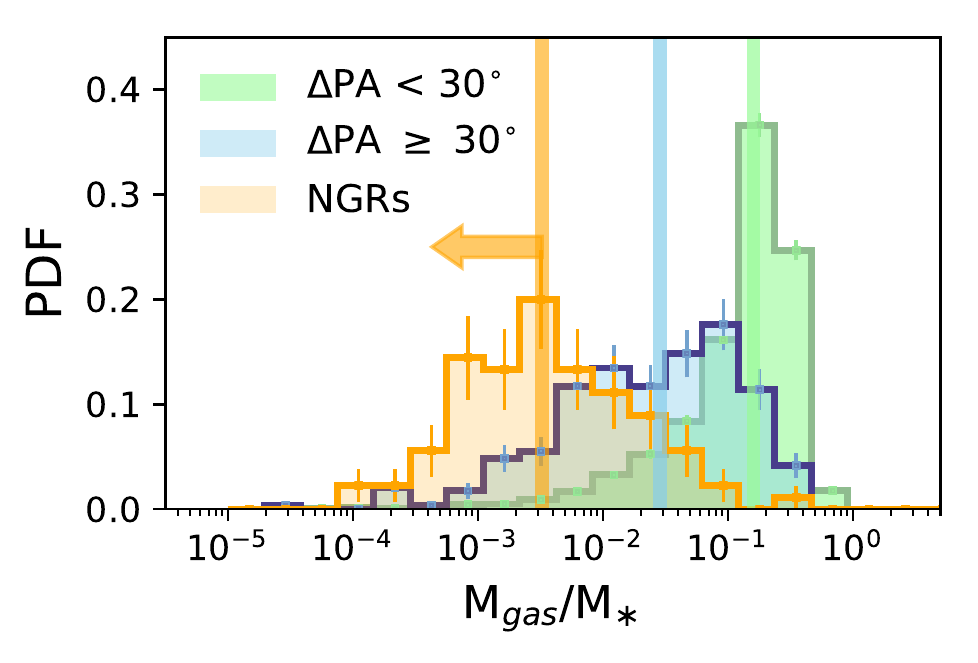}
    \caption{Probability density distributions of gas mass fraction, $(M_{gas}/M_{\ast})$ for aligned galaxies ($\Delta$PA < 30$^{\circ}$) shown in green, those with high misalignment ($\Delta$PA > 30$^{\circ}$) in light blue and NGRs in orange. Each histogram is given with Poisson errors on each bin. The vertical lines denote the corresponding distribution's median. The majority of NGRs do not have detectable gas masses and therefore the distribution shown should be considered as upper bound.}
    \label{fig:delPA_gasM}
\end{figure}

In Figure \ref{fig:delPA_gasM} we show the distribution of gas masses for the three populations. We see a clear trend of lower gas mass going from kinematically aligned galaxies to misaligned galaxies to NGRs. We note that the majority ($\sim$80\%) of NGRs do not contain enough gas to have a measured gas mass from the routine of Pipe3D, so the distribution shown is a hard upper limit on the gas that these galaxies contain. We note that these trends remain qualitatively the same when considering the distributions for ETGs, S0-Sas and Sb-Sds individually.

The similarity in stellar angular momentum between the NGRs and kinematically misaligned galaxies could indicate that they are from the same evolutionary sequence. A key component in decoupling star-gas rotation in simulations is a significant gas loss followed later by the accretion of material with misaligned angular momentum \citep[][]{vdvoort2015, starkenburg+19}. This gas loss can happen due to interactions from neighbouring galaxies which strips gas or through ejection due to black hole feedback.

In \citet{duckworth2019}, it was shown that kinematic decoupling shows little relationship with distance to filamentary structure. This could point to stripped/ejected material being re-accreted as a potential source of misalignment between star and gas rotation. Some NGRs could therefore represent an earlier timestamp before this material is re-accreted. Not all NGRs would necessarily re-accrete gas, meaning that some would remain quenched (and hence would not become misaligned in the future) potentially explaining the differences we see in stellar mass distributions of NGRs and misaligned. In this scenario, it would suggest that re-accretion of new material does not significantly alter the stellar angular momentum content going from NGRs to misaligned.

\subsection{Morphology}
\begin{figure}
	\includegraphics[width=\linewidth]{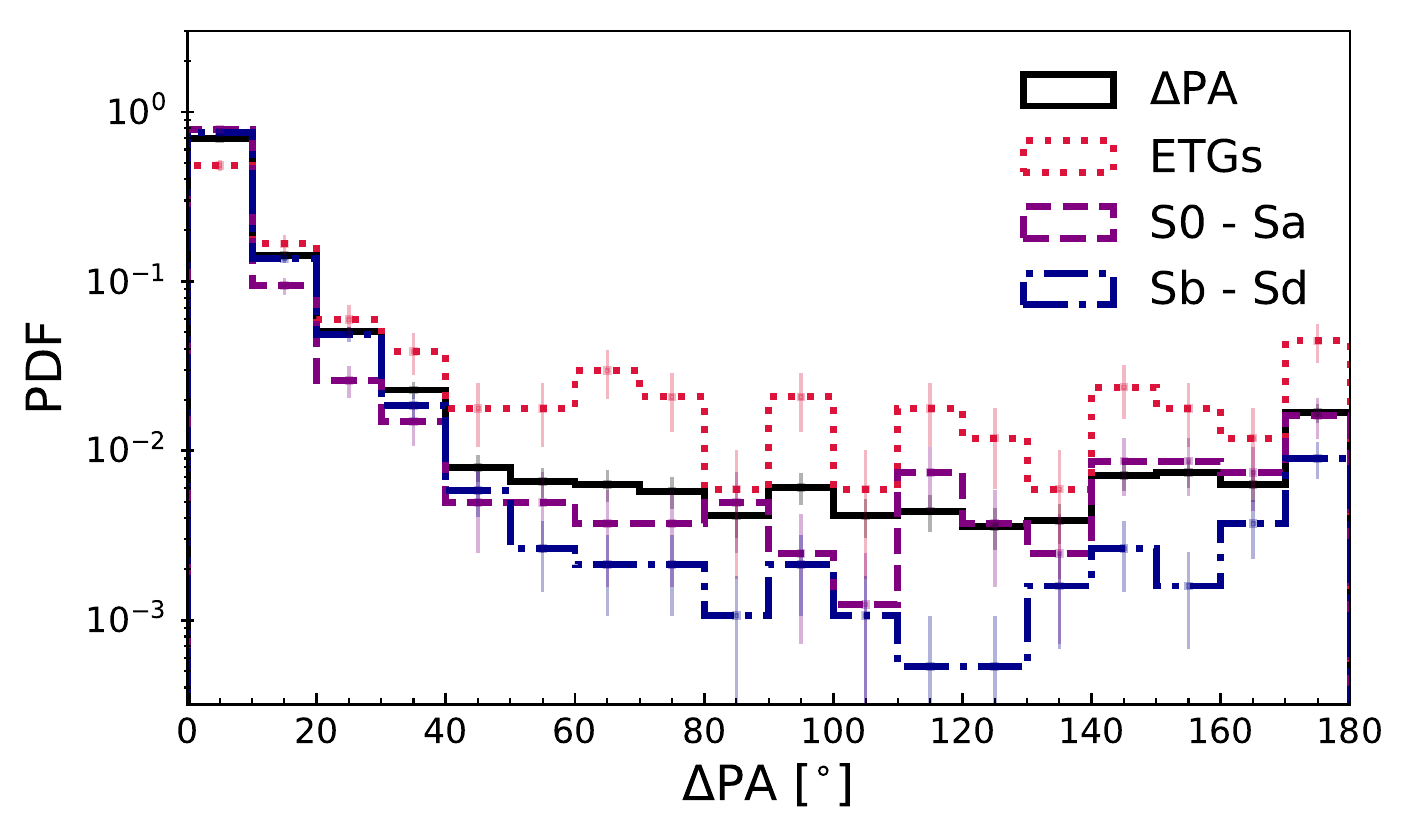}
    \caption{Probability density distributions of kinematic misalignment as defined by $\Delta$PA split on morphology. The probability density distribution is normalised to 1 and shown in log scale. Distributions for the total population, ETGs, S0/Sa and Sb-Sds are shown by black solid, dotted red, dashed purple and dot-dashed blue lines respectively. Earlier type galaxies are more likely to be misaligned than later type galaxies.}
    \label{fig:morph_PA}
\end{figure}

We now sub-divide the total population by morphology into ETGs, S0-Sas and Sb-Sds as defined in \S\ref{sec:morph_def}. Figure \ref{fig:morph_PA}, shows the distributions for each category. We find that for all morphological types, galaxies are most commonly aligned with strong peaks below $\Delta$PA $\sim 30^{\circ}$. ETGs show a flatter distribution than their later counterparts, as the most likely to exhibit misalignment. LTGs show deeper drop-offs above $\Delta$PA $\sim 40^{\circ}$, with a boost around $\Delta$PA = 180$^{\circ}$, seen most strongly for the Sb-Sds. We quantify the overall misalignment fractions in the first column of Table \ref{tab:mega_table}. Our errors are estimated by binomial counting errors so that $\sigma = \sqrt{p(1-p) / M}$ where $p = N/M$ with $N$ being the number of misaligned galaxies and $M$ the total number of galaxies for the category.

This morphological difference in misalignment is likely a result of several factors. Gas rich LTGs have typically higher specific angular momentum, and hence, require a higher magnitude gas inflow/outflow with different angular momentum to disrupt rotation and create misalignment. Conversely, ETGs are more dispersion dominated and gas poor enabling smaller gas in-flows (or outflows) to create a kinematic misalignment. 

These results are reasonably consistent with previous findings of 36$\pm$5\% (of 260 galaxies) of ETGs that are misaligned in ATLAS\textsuperscript{3D} and in SAMI (45$\pm$6\% of 36 pure ellipticals, 5$\pm$1\% in 221 pure late spirals) \citep[][]{davis2011, bryant2019}. We note that our ETG misalignment fraction ($\sim$28\%) is lower than these previous findings and holds a slight tension with \citet{bryant2019}. Possible reasons for the differences may be due to morphology definition, stellar mass distribution or simply sample size. We note that enforcing stricter thresholds for morphology classifications doesn't change our misaligned fractions pointing to a likely difference in mass distributions or our increased sample size. 

\begin{table*}
\begin{tabular}{lllll}
\hline
        &  & All & Centrals & Satellites \\
\hline
All galaxies & $\Delta$PA defined &  3798 &  2185 &  1007 \\
& $\Delta$PA $\geq 30^{\circ}$ &  420 (11.1$\pm$0.5\%) &  251 (11.5$\pm$0.7\%) &  102 (10.1$\pm$1.0\%) \\
& NGR & 742 &  334 &  324 \\

ETGs & $\Delta$PA defined & 301 & 204 & 97 \\
& $\Delta$PA $\geq 30^{\circ}$ & 84 (27.9$\pm$2.6\%) & 60 (29.4$\pm$3.2\%) & 24 (24.7$\pm$4.4\%)  \\
& NGR & 231 & 140 & 91 \\

S0 - Sas & $\Delta$PA defined & 677 & 483 & 194 \\
& $\Delta$PA $\geq 30^{\circ}$ &  66 (9.7$\pm$1.1\%) & 49 (10.1$\pm$1.4\%) & 17 (8.8$\pm$2.0\%) \\
& NGR & 100 & 44 & 56 \\

Sb - Sds & $\Delta$PA defined & 1634 & 1112 & 522 \\
& $\Delta$PA $\geq 30^{\circ}$ & 88 (5.4$\pm$0.6\%) & 58 (5.2$\pm$0.7\%) & 30 (5.7$\pm$1.0\%) \\
& NGR & 107 & 32 & 75 \\

\end{tabular}
\caption{Total number of galaxies used in this study for each of $\Delta$PA defined sample, of those that are kinematically misaligned and those that have well defined stellar rotation but incoherent gas (NGR). These are defined for both splitting on morphology (rows) and group membership (columns). For those that are kinematically misaligned ($\Delta$PA $\geq 30^{\circ}$), the percentage with respect to all those with $\Delta$PA measurements for the sub-category is shown. The uncertainties quoted are binomial counting errors.}
\label{tab:mega_table}
\end{table*}

The boost in the PDF around 180$^{\circ}$ of Figure \ref{fig:morph_PA} suggests that near counter-rotation is a stable state for galaxies. This is seen most prominently in Sb-Sds with a clear upwards trend in the PDF from $\sim$140$^{\circ}$. A possible explanation is that these rotation dominated galaxies host strong stellar torques, which act to realign gas on much faster timescales than in ETGs. Counter-rotators, however, remain stable and hence contribute proportionally higher to the misaligned distribution, as those at intermediate misalignments settle towards alignment. 

Interestingly galaxies that exhibit near-counter rotation ($\Delta$PA $\geq$ 150$^{\circ}$) have similar stellar angular momentum to the general misaligned population ($\Delta$PA $\geq$ 30$^{\circ}$), significantly lower than the aligned counterparts. This holds true for all morphologies. \citet{chen2016} previously highlighted the boost in star formation in central regions for counter-rotating LTGs. As suggested, this could be a natural result of cancellation of angular momentum leading to increased in-flows to central regions. Our finding of lowered angular momentum in the counter-rotators (with respect to the co-rotators) supports this claim.

\begin{figure}
	\includegraphics[width=\linewidth]{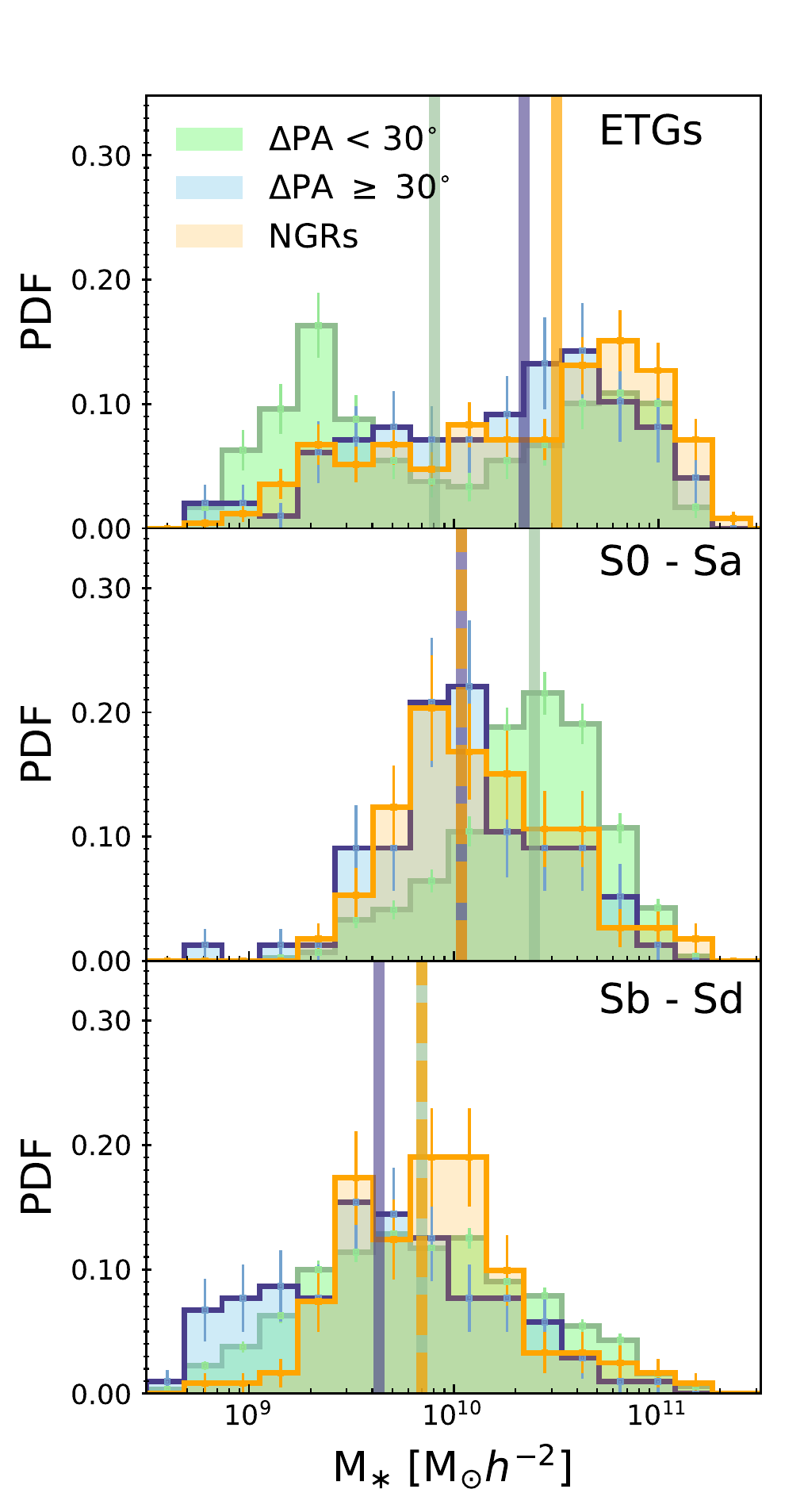}
    \caption{Probability density distributions of stellar mass, $(M_{\ast}/M_{\odot})$ for aligned galaxies ($\Delta$PA < 30$^{\circ}$, misaligned galaxies ($\Delta$PA > 30$^{\circ}$) and NGRs for ETGs, S0-Sas and Sb-Sds (top to bottom). In each panel the aligned/misaligned are shown with solid lines with the aligned in the darker shade. NGRS are shown by dot-dashed lines. Each histogram is given with Poisson errors on each bin. The vertical lines denote the corresponding distribution's median. For ETGs, aligned galaxies are less massive than the misaligned sample. This trend, however, reverses for S0-Sas and Sb-Sds.}
    \label{fig:morph_stelM}
\end{figure}

Due to the relationship between stellar mass, morphology and specific angular momentum \citep[e.g.][]{cortese2016}, it might be expected that misaligned galaxies should be at higher stellar mass due to their lower $\lambda_{R}$ with respect to the aligned \citep[see also;][]{bryant2019}. Surprisingly for the overall population we see little difference, however, splitting on morphology as shown in Figure \ref{fig:morph_stelM} reveals individual trends. Misaligned ETGs (and NGRs) are more massive than the aligned counterparts most likely indicative that misaligned galaxies have had richer merger histories. The opposite trends are seen for both S0-Sas and Sb-Sds with kinematically aligned galaxies being of typically higher mass than the misaligned. This could be indicative that the pathways leading to misalignment are different as a function of morphology.


\subsection{Group membership}
\begin{figure*}
	\includegraphics[width=\linewidth]{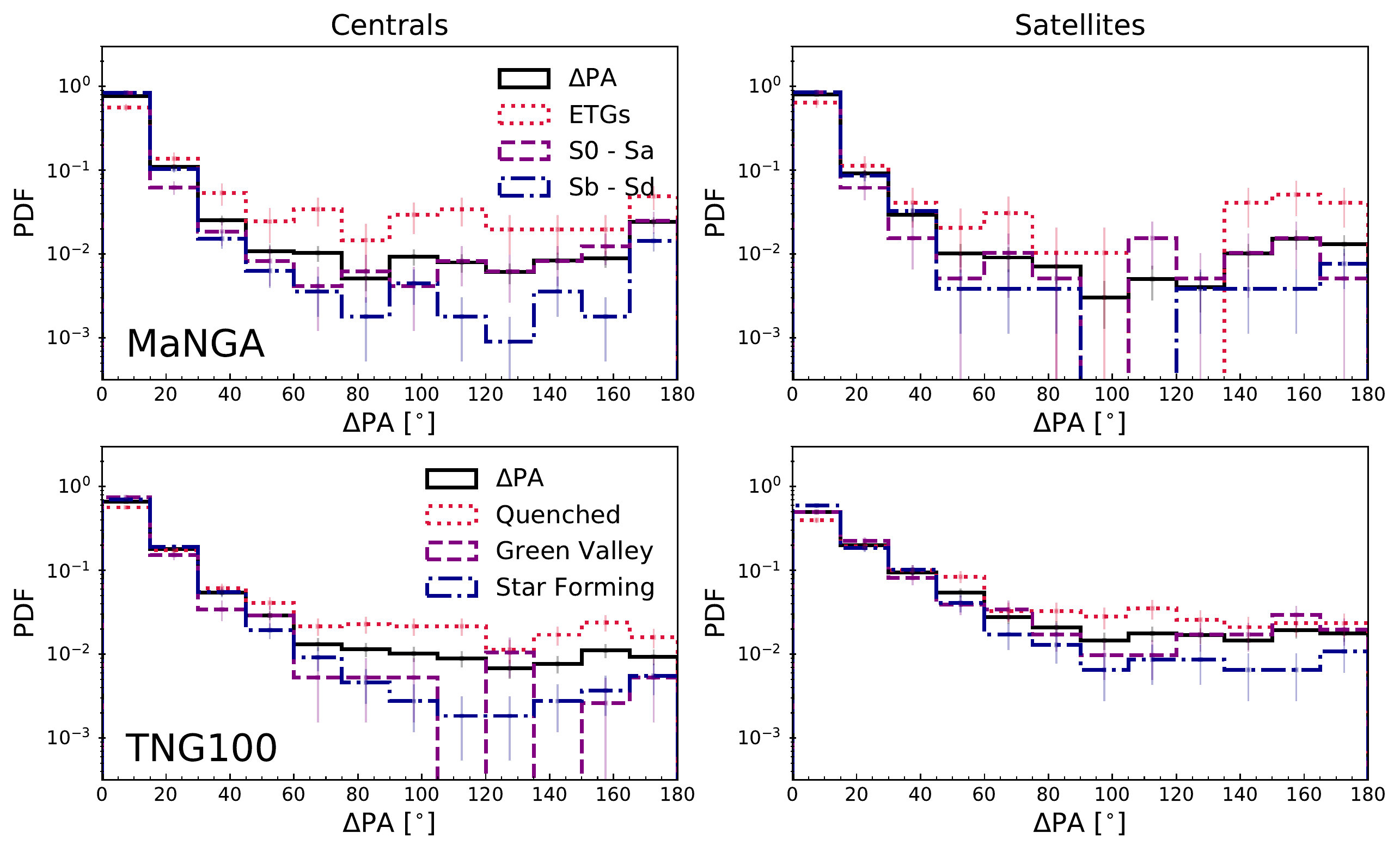}
    \caption{Same as Figure \ref{fig:morph_PA}, however split by group membership into centrals (left) and satellites (right). The top panel shows for the MaNGA sample and the bottom shows for the mock sample in TNG100. Morphology for TNG100 is categorised by the deviation of the galaxy's star formation away from the main sequence of galaxies in the whole of TNG100 (see \S\ref{sec:tng_results})}
    \label{fig:group_morph_PA}
\end{figure*}

Group membership is important for dictating the evolution of a galaxy and hence we now sub-divide our population into centrals and satellites as described in \S\ref{sec:group_def}. Figure \ref{fig:group_morph_PA} (top panels) shows the $\Delta$PA distributions as in Figure \ref{fig:morph_PA}, but now split into centrals and satellites. Qualitatively the morphological trends remain however Table \ref{tab:mega_table} reveals that centrals (29.4$\pm$3.2\%) are slightly more likely to be misaligned than satellites (24.7$\pm$4.4\%) for ETGs. This is also potentially seen for the S0-Sbs (10.1$\pm$1.4\% for centrals vs 8.8$\pm$2.0\% for satellites), however we note that both fractions are within each other's errorbars.

\begin{figure*}
	\includegraphics[width=0.75\linewidth]{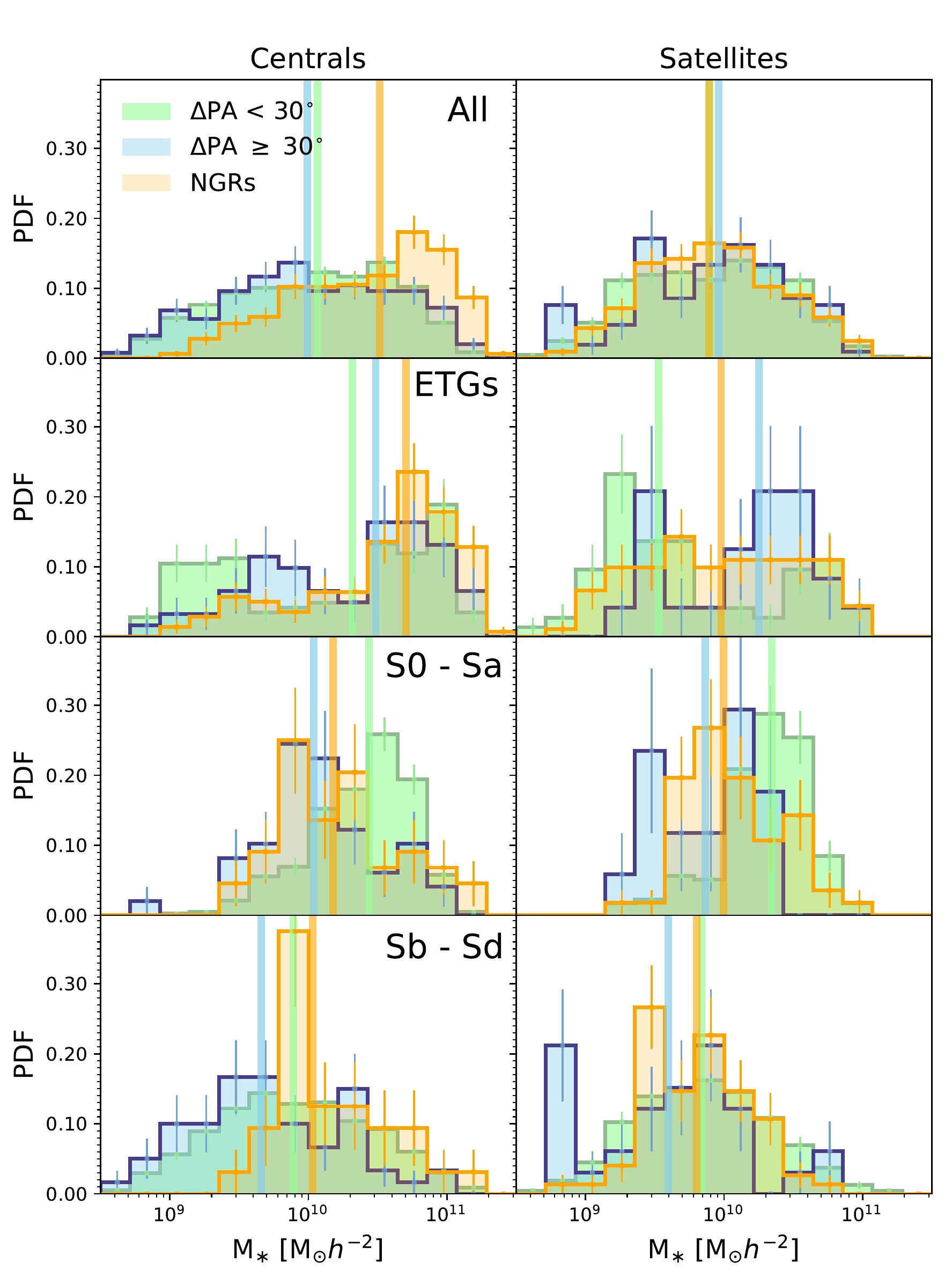}
    \caption{Same as Figure \ref{fig:morph_stelM}, however split by group membership into centrals (left) and satellites (right). Additionally the distributions for the overall central and satellite populations is shown in the top row. We see that for ETGs there is a strong difference in mass between aligned and misaligned satellites. This trend is reversed for S0/Sa and Sb/Sd satellites. These trends are also seen for centrals, however, typically to a lesser degree.}
    \label{fig:group_morph_stelM}
\end{figure*}

Figure \ref{fig:group_morph_stelM} shows the stellar mass distribution for our samples but now additionally split into centrals and satellites. Again we find the same qualitative trends for both centrals and satellites; i.e. misaligned ETGs are more massive than their aligned counterparts whereas misaligned S0-Sas and Sb-Sds are less massive than their aligned counterparts.

\section{Misaligned galaxies in TNG100} \label{sec:tng_results}
In this section, we utilise the mock sample created in TNG100 to interpret the properties of kinematically misaligned galaxies in MaNGA and isolate the driving mechanisms. 

We divide our mock MaNGA sample based on the instantaneous star formation rate within twice the stellar half mass radius of a given galaxy with respect to all galaxies in TNG100. The star forming main sequence for all galaxies is found by fitting a power law as a function of mass. A galaxy is then flagged into one of three categories; star forming, green valley or quenched depending on its deviation above or below the main sequence \citep{pillepich2019}. The selected deviations from the main sequence are as follows; star forming galaxy: $\Delta \log_{10}(SFR) > −0.5$, green valley galaxy: $-1.0 < \Delta \log_{10}(SFR) < -0.5$ and quenched galaxy: $\Delta \log_{10}(SFR) <= -1.0$.

The bottom panel of Figure \ref{fig:group_morph_PA}, shows the $\Delta$PA distribution for the TNG100 sample split into centrals and satellites. Comparing to the observational sample in the top panel of Figure \ref{fig:group_morph_PA}, the morphological trends remain qualitatively the same with quenched/ETGs (star forming/LTGs) more likely to be misaligned (aligned).

Our choice to compare populations split on visual morphology in observations to sSFR in simulations is one of necessity. The aim of this work is to explore the relationship of visual morphology with decoupled rotation. Unfortunately we don't currently have the equivalent classifications in IllustrisTNG100, so use an appropriate proxy. In future work, we will look at the relationship between observations and simulations using machine learning classifications of morphology, however, in the following subsections we follow the evolutionary history of the mock sample split by sSFR.

\subsection{Angular momentum}
In this sub-section, we consider the angular momentum content of our TNG100 mock sample back to $z=1$ for stars, gas and dark matter individually. Angular momentum for our TNG100 galaxies is defined by the intrinsic specific angular momentum of their particles/cells:
\begin{equation}
j_{k} = \frac{1}{\sum_{n} m^{(n)}} \sum_{n} m^{(n)}\boldsymbol{x}^{(n)} \times \boldsymbol{v}^{(n)}
\end{equation}
where $\boldsymbol{v}^{(n)}$ is the velocity of each particle relative to the centre of mass for the galaxy. $\boldsymbol{x}^{(n)}$ is the position of a given particle with respect to the position of the most gravitationally bound particle in the galaxy. We choose this definition since the centre of mass velocity can be biased by structure at large radii and hence may spuriously not represent the true rotational centre. $k$ is the particle/cell type referring to either stars, gas or dark matter. For stars and gas this is calculated within the 3D radius equivalent to sky coverage assigned by the mock IFU observation. Dark matter is calculated for all particles assigned to the subhalo by the subfind algorithm. 

\begin{figure*}
	\includegraphics[width=\linewidth]{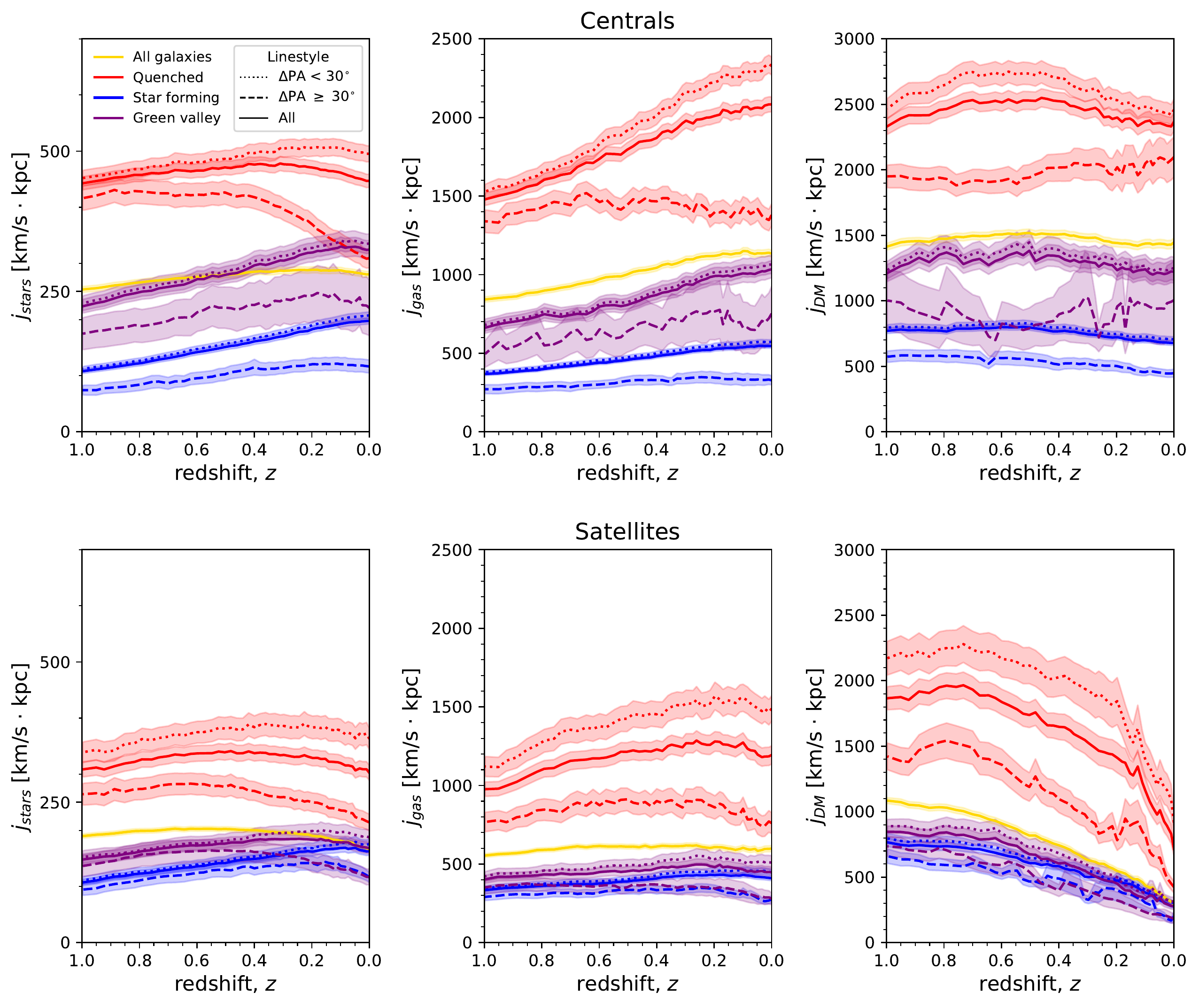}
    \caption{Specific angular momentum evolution from $z = 1$ calculated from star, gas and DM particles (left to right). The angular momentum is calculated for all star and gas particles/cells within the 3D radius assigned by the mock IFU observation, whereas DM is found from all particles associated to the subhalo. The evolution is taken as the median at each timestep for all galaxies of that category with errorbars showing the standard error. The top (bottom) row shows the evolution for central (satellite) galaxies. Each panel displays the evolution split into morphologies; quenched (red), green valley (purple) and star forming (blue) and also $\Delta$PA $< 30^{\circ}$ (dotted) and $> 30^{\circ}$ (dashed). Kinematically misaligned galaxies selected at $z=0$ have notably lower specific angular momentum for all of stars, gas and dark matter.}
    \label{fig:sJ_evo}
\end{figure*}

Figure \ref{fig:sJ_evo} shows the specific angular momentum evolution from $z=1$ for all components split on group membership and morphology. We see that similar to the observational sample (see \S\ref{sec:manga_total_pop}), misaligned galaxies in simulations are significantly lower angular momentum than their aligned counterparts at $z=0$ for the stellar components. This is reflected in all components (stars, gas, DM) to various degrees for all morphologies and central/satellite definition. Interestingly, while misalignment between stars and gas may itself be a transient property, those misaligned at $z = 0$ reside in dark matter haloes with \textit{fundamentally lower angular momentum} which persists to at least $z = 1$. 

\begin{figure*}
	\includegraphics[width=\linewidth]{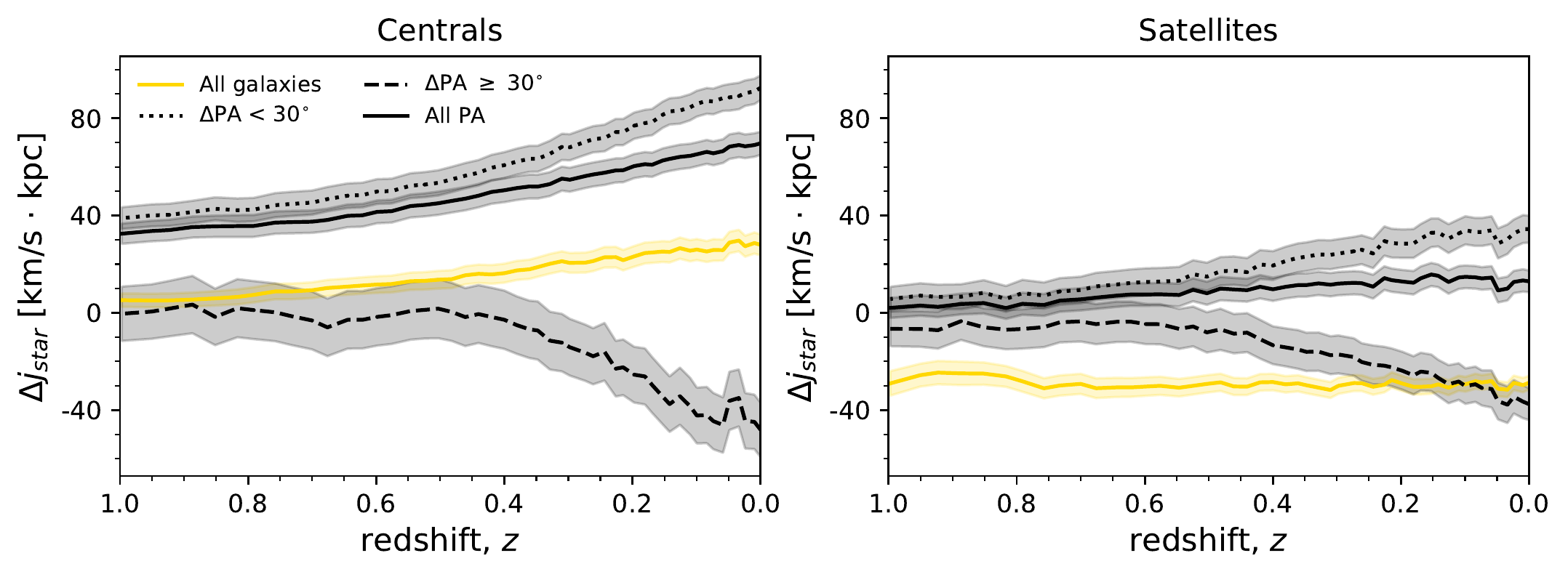}
    \caption{The specific angular momentum residuals from $z=1$ for all star particles within the 3D radius assigned by the mock IFU observation. The residual is calculated as the deviation away from the expectation for a galaxy of that mass at each snapshot. The evolution of the residual is taken as the median at each timestep for all galaxies of that category with errorbars showing the standard error. The right (left) panel shows the evolution for central (satellite) galaxies. Each panel displays the evolution for all galaxies (yellow), of which have a defined $\Delta$PA (black solid), aligned galaxies $\Delta$PA $< 30^{\circ}$ (black dotted) and misaligned $> 30^{\circ}$ (black dashed). We see that the difference in angular momentum between aligned and misaligned galaxies is not due to differences in mass. In addition we note a marked deviation of misaligned galaxies to even lower angular momentum in recent times.}
    \label{fig:sJ_evo_residual}
\end{figure*}

We note that particle based calculations of specific angular momentum scales with the number of particles. This results in more massive galaxies having higher $j_{i}$ and further, quenched galaxies (that are typically more massive) having higher $j_{i}$ than their later type counterparts. While there is only a small difference in-between the mass distributions of our aligned and misaligned samples, to ensure our signal is not simply driven by mass we calculate the residuals of $j_{star}$ with respect to a typical galaxy of that mass. The residuals, $\Delta j_{star}$ are calculated by fitting a polynomial to the distribution of $j_{star}$ vs $M_{\ast}$ for the galaxies (all mock observations, regardless if $\Delta$PA is well defined) at each snapshot. $\Delta j_{star}$, is then defined as the deviation of a given galaxy away from the expectation of the fitted line at that mass. Since the trends are qualitatively consistent regardless of morphology, Figure \ref{fig:sJ_evo_residual} shows the specific angular momentum residuals for the total population. For completeness we also include the expectation for all matched galaxies regardless of if $\Delta$PA is well defined. Misaligned galaxies ($\Delta$PA $\geq 30^{\circ}$) for both centrals and satellites show intrinsically lower $\Delta j_{star}$ with respect to the total population at a given mass, indicative that it is not an effect due to mass. In addition, there is a relative evolution where $\Delta j_{stars}$ diverges from all galaxies at $z \sim 0.5$ so that misaligned galaxies have even lower stellar angular momentum with respect to the aligned galaxies in recent times.

To conclude this section we now consider the directional offsets between the angular momentum vectors of the stars, gas and dark matter. These are calculated from:
\begin{equation} \label{eq:alpha}
    \alpha_{3D} = \text{arccos} \left( \frac{\boldsymbol{j_{i}} \cdot \boldsymbol{j_{j}}}{\left| \boldsymbol{j_{i}} \right| \left| \boldsymbol{j_{j}} \right|} \right),
\end{equation}
where $i, j$ refer to either stars, gas or dark matter. As for the magnitudes of angular momentum, the star and gas vectors are calculated within a 3D radius set to that of the IFU footprint and the dark matter vector is calculated for all particles assigned to the subhalo by subfind. A comparison between calculating star-gas decoupling from angular momentum and from PA fitting can be found in Appendix \ref{sec:PA_test}. Figure \ref{fig:3D_alpha_evo} shows the evolution of the 3D offsets between each of stars, gas and DM respectively. 

As expected splitting our sample on $\Delta$PA results in significantly higher $\alpha_{STARS - GAS}$ at $z = 0$ for the misaligned galaxies found in the MaNGA observations. This is also typically correlated, albeit less strongly, with larger $\alpha_{STARS - DM}$ and $\alpha_{GAS - DM}$ at $z=0$. This is indicative that a decoupling between stars and gas is often mirrored by a decoupling between the rotations of stars and DM. We also plot the average decoupling for all galaxies (all that are matched to MPL-8) between all components. In the middle panel, we see that $\alpha_{STARS - DM} \sim 50^{\circ}$ on average for all galaxies (gold line) with a slight redshift evolution which is roughly consistent with previous work \citep[e.g.][]{chisari+17}. We note that our choice to consider the direction of the star and gas rotation within the observational footprint is typically far smaller than the overall DM halo, and hence, may lead to slightly higher typical misalignments between baryonic and DM components.

Working back from $z=0$, we note that $\alpha_{STARS - GAS}$ for the aligned and misaligned samples (selected at $z=0$) converges in the majority of cases before $z=1$. This indicates the transient nature of misalignment. This is in stark contrast to the magnitude of angular momentum for individual components (stars, gas, DM) which show a persistent difference in magnitude between aligned and misaligned objects (selected at $z=0$) going back past at least $z=1$. 

\begin{figure*}
	\includegraphics[width=\linewidth]{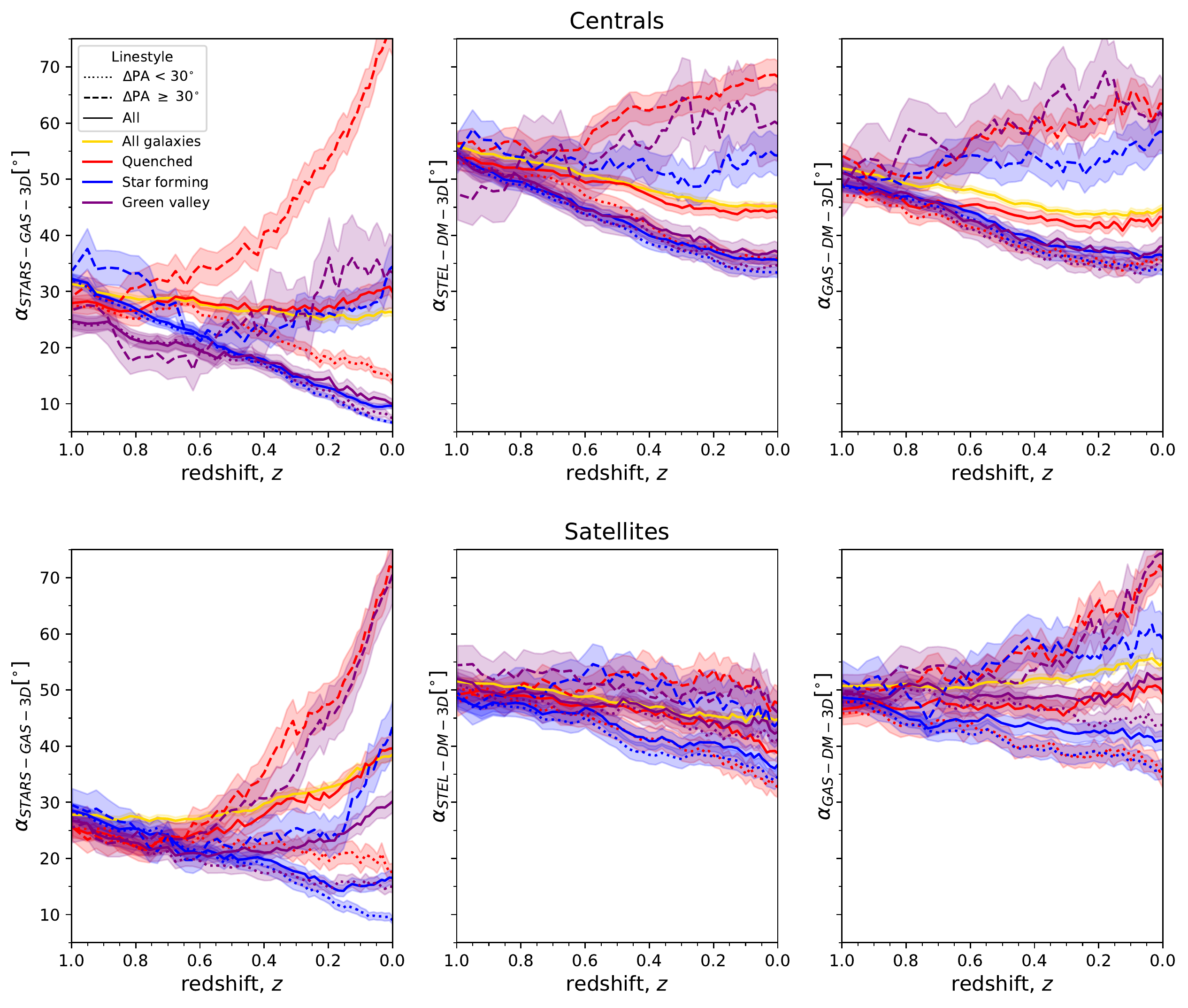}
    \caption{Evolution of the 3D offset (in degrees) between the principal spin axes of; stars and gas (left), stars and dark matter (middle) and gas and dark matter (right) from $z=1$. The evolution is taken as the median at each timestep for all galaxies of that category with errorbars showing the standard error. The top (bottom) row shows the evolution for central (satellite) galaxies. Each panel displays the evolution split into morphologies; quenched (red), green valley (purple) and star forming (blue) and also $\Delta$PA $< 30^{\circ}$ (dotted) and $\geq 30^{\circ}$ (dashed).}
    \label{fig:3D_alpha_evo}
\end{figure*}

\section{Discussion} \label{sec:discussion}
In the previous sections we have demonstrated the relationship of kinematic misalignment with morphology, stellar angular momentum and dark matter halo spin. In the following we put our results in context and highlight the potential of using the decoupling of star-gas rotation to identify underlying properties of a galaxy. 

We note the close relationship of our findings of our samples with respect to the work of \citet{starkenburg+19}. They investigate the origin of star-gas decoupling (in this instance $ > 90^{\circ}$) using individual case studies of low mass galaxies (i.e. $2 \times 10^{9} < M_{\ast} < 5 \times 10^{10}$) in the original Illustris simulation. Despite extending the mass range and only considering the ensemble average for aligned and misaligned galaxies split at $\Delta$PA$= 30^{\circ}$, we still find the same qualitative trends of lower angular momentum and lower gas mass fractions for misaligned galaxies (in comparison to aligned). 

While outside the scope of this work, we note that their estimation of relaxation timescales (i.e. until realignment of rotation axes) is of the order Gigayears. This appears to be roughly comparable to toy-model estimates \citep[see;][albeit for ETGs]{davis2016}. Here we also demonstrate the transient nature of star-gas decoupling (left panels, Figure \ref{fig:3D_alpha_evo}). Working back from $z=0$, we note that $\alpha_{STARS - GAS}$ for the aligned and misaligned samples (at $z=0$) converges in the majority of cases before $z=1$. Since we are only considering the ensemble average for misalignment selected at $z=0$, we cannot comment on the timescales of misalignment here since the average may include several events that decouple the rotation.

In contrast, we see that the magnitude of specific angular momentum for stars, gas and DM for misaligned objects (at $z=0$) remains fundamentally lower going to at least $z=1$. This suggests that while star-gas decoupling at $z=0$ is a transient property, \textit{its likelihood is correlated with the angular momentum content of the halo at early times}. In part, the correlation must be driven by the lower angular momentum content of the stellar component. This inherently leads to longer relaxation timescales (i.e. longer star-gas decoupling) due to weaker stellar torques acting on the misaligned gas component and hence a higher likelihood of being misaligned at $z=0$

We also note TNG100s ability to not only reproduce a reasonable distribution of $\Delta$PAs with respect to the MaNGA sample (Figure \ref{fig:total_pa_dist}) once accounting for variances in mass between the $\Delta$PA defined samples in MaNGA and TNG100 but also reproducing the strong trends with morphology found in observations (Figure \ref{fig:group_morph_PA}). 

Whether the trigger of misalignment is internal or external, it appears to be clearly linked to a lowered gas mass (Figure \ref{fig:delPA_gasM}). In future work we will use our observational and simulated samples to break down the prevalence of driving factors. 

\section{Conclusion} \label{sec:conclusion}
In this work, we introduce a catalogue of $\sim$4500 galaxies from the MaNGA survey in order to establish the prevalence of misalignment as a function of morphology and angular momentum. We also construct a mock MaNGA sample in IllustrisTNG100 to determine the time evolution of angular momentum in star-gas decoupled galaxies and their relationship with halo spin. Our conclusions are as follows:
\begin{enumerate}

    \item (MaNGA) The prevalence of kinematic misalignment (i.e. where rotational axes of stars and gas are offset by $> 30^{\circ}$) is strongly morphological dependent with ETGs having $\sim$28\% exhibiting misalignment which decreases to $\sim$5\% for Sb-Sds.
    
    \item (MaNGA) For all morphologies this misalignment is related to a lowered stellar angular momentum and also a lowered gas mass. We note that misaligned galaxies have similar stellar angular momentum to those do not have coherently rotating gas (those with large gas depletion fall into this category). This could be indicative that galaxies without coherent gas rotation and kinematically misaligned galaxies are different timesteps in the same evolutionary sequence. As noted in simulations \citep[][]{vdvoort2015, starkenburg+19}, a key component in decoupling star-gas rotation is a significant gas loss followed by accretion of new gas with misaligned angular momentum. In this scenario, NGRs could represent an earlier timestamp before a future re-accretion of gas. This would indicate that the stellar angular momentum is disrupted prior to accretion of new material. 
    
    \item (MaNGA) We find that the misalignment fraction is also dependent on group membership. For ETGs and S0-Sas, central galaxies are more likely to exhibit misalignment than satellites. For Sb-Sds this trend reverses.
    
    \item (MaNGA) We find that counter-rotation (i.e. rotational axes of stars and gas are offset by $> 150^{\circ}$) is a stable state for galaxies of all morphologies shown by a boost in the PDF (Figure \ref{fig:morph_PA}). Similar to the total misaligned population, counter-rotators have distinctly lower angular momentum than their aligned counterparts. 
    
    \item (IllustrisTNG100) We find that a mock MaNGA like sample constructed from cosmological scale hydrodynamical simulation IllustrisTNG100 reproduces the observed trends of decoupling with morphology and stellar angular momentum at $z=0$.
    
    \item (IllustrisTNG100) We find that decoupled galaxies reside in dark matter haloes with lower spin going back past $z=1$. Despite the decoupling between gas and stars being inherently transient in nature, it is also associated with a decoupling of both stars and gas with respect to dark matter. This demonstrates the inherent link of decoupling, not only to present day stellar angular momentum, but to lower spin haloes at $z=1$. 

\end{enumerate}
In future work we will use these observational and simulated samples to further investigate the different drivers of misalignment as a function of morphology. In particular we will explore the relationship with BH luminosity and feedback, and, investigate the typical timescales of misalignment and merger rates in ETGs. 

\bibliographystyle{mnras}
\bibliography{biblio.tex}


\section*{Acknowledgements}
CD acknowledges support from the Science and Technology Funding Council (STFC) via an PhD studentship (grant number ST/N504427/1) and from the Simons Foundation through its support of the Flatiron Institute. 

We thank the IllustrisTNG team for help and data access. While this work was conducted using the public data release, it would have not been possible without their help and use of the IllustrisTNG private data for the purposes of other research works. 

SDSS-IV is managed by the Astrophysical Research Consortium for the Participating Institutions of the SDSS Collaboration including the Brazilian Participation Group, the Carnegie Institution for Science, Carnegie Mellon University, the Chilean Participation Group, the French Participation Group, Harvard-Smithsonian Center for Astrophysics, Instituto de Astrof\'isica de Canarias, The Johns Hopkins University, Kavli Institute for the Physics and Mathematics of the Universe (IPMU) / University of Tokyo, Lawrence Berkeley National Laboratory, Leibniz Institut f\"ur Astrophysik Potsdam (AIP), Max-Planck-Institut f\"ur Astronomie (MPIA Heidelberg), Max-Planck-Institut f\"ur Astrophysik (MPA Garching), Max-Planck-Institut f\"ur Extraterrestrische Physik (MPE), National Astronomical Observatory of China, New Mexico State University, New York University, University of Notre Dame, Observat\'orio Nacional / MCTI, The Ohio State University, Pennsylvania State University, Shanghai Astronomical Observatory, United Kingdom Participation Group, Universidad Nacional Aut\'onoma de M\'exico, University of Arizona, University of Colorado Boulder, University of Oxford, University of Portsmouth, University of Utah, University of Virginia, University of Washington, University of Wisconsin, Vanderbilt University, and Yale University.
\appendix

\section{PA vs absolute offset} \label{sec:PA_test}
\begin{figure}
	\includegraphics[width=\linewidth]{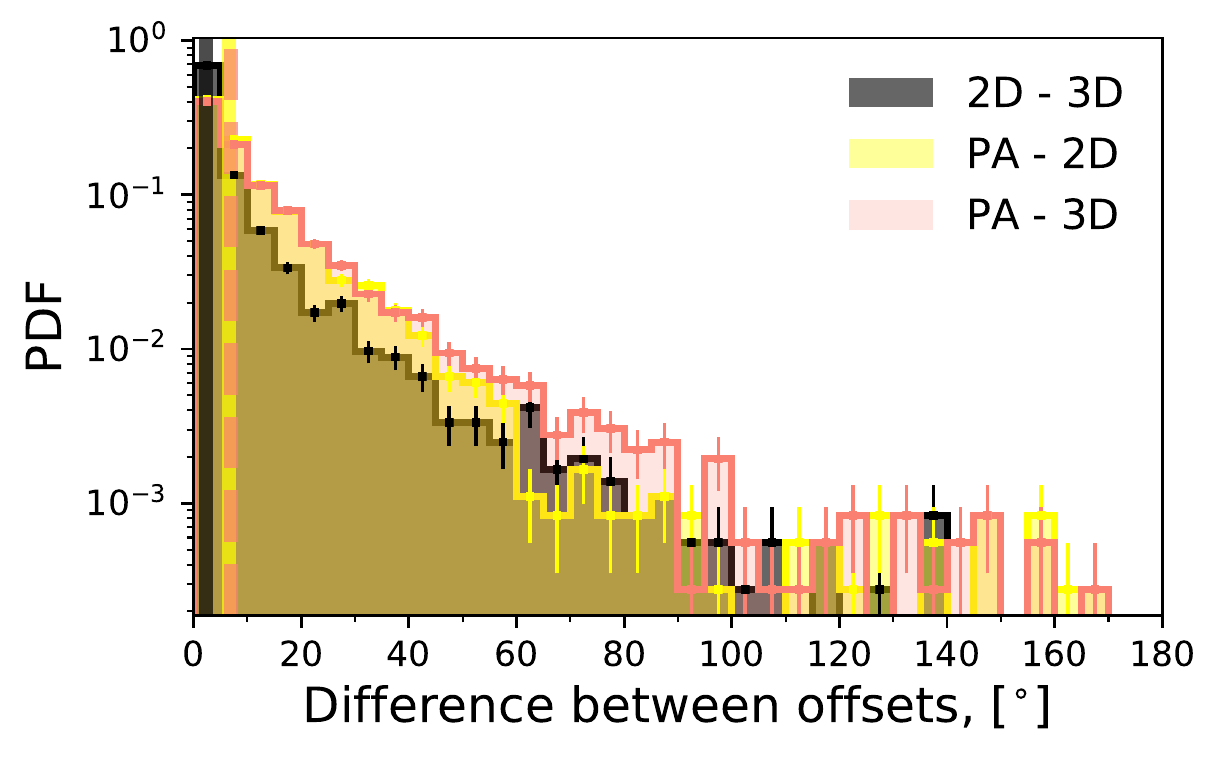}
    \caption{Probability density distribution of the difference between various measures of the star-gas rotational angle offset. The difference between the 3D angular momentum vectors and projection in 2D is shown (black), $\Delta$PA and 2D (yellow) and $\Delta$PA and 3D (blue).}
    \label{fig:PA_residual}
\end{figure}

A key assumption of this work is the ability for the projected $\Delta$PA to be a reliable estimator of the actual 3D offset between star and gas rotation axes. Figure \ref{fig:PA_residual} shows the distributions of the difference between $\Delta$PA and the 2D and 3D offsets between the angular momentum principal axes of stars and gas. See equation \ref{eq:alpha} for calculation of the 3D offset; the 2D equivalent is simply a projection of this onto the XY plane. $\Delta$PA is a reasonable measure of the true 3D offset which can be modelled as a Gaussian centred on 0$^{\circ}$ with a standard deviation of 17.6$^{\circ}$ (green dotted line). The deviation of the 2D projection from the true 3D offset (black line) has a standard deviation of 13$^{\circ}$, demonstrating that the variation is both due to projection and the noise associated with observations. Additionally, we note the different particle selection for the two measures which may drive slight differences. While the 2D/3D offsets and $\Delta$PA are measured in a footprint with the same radius, the offsets are only defined for particles within a 3D sphere of this radius, where $\Delta$PA is defined for all particles along the line of sight enclosed by the sky footprint. 

\bsp	
\label{lastpage}
\end{document}